\documentclass[twocolumn,pra,showpacs,superscriptaddress]{revtex4-1}
\pdfoutput=1
\usepackage{amsmath, amssymb}
\usepackage{pstricks}
\usepackage{graphicx,color}
\usepackage{hyperref}
\hypersetup{colorlinks,linkcolor=blue,citecolor=blue,urlcolor=blue}

\makeatletter
\let\cat@comma@active\@empty
\makeatother

\begin{document}

\title{A coordinate Bethe ansatz approach to the calculation of equilibrium and nonequilibrium correlations of the one-dimensional Bose gas}

\author{Jan C Zill}
\email{j.zill@uq.edu.au}
\affiliation{School of Mathematics and Physics, The University of Queensland, Brisbane QLD 4072, Australia}

\author{Tod M Wright}
\affiliation{School of Mathematics and Physics, The University of Queensland, Brisbane QLD 4072, Australia}

\author{Kar{\'e}n V Kheruntsyan}
\affiliation{School of Mathematics and Physics, The University of Queensland, Brisbane QLD 4072, Australia}

\author{Thomas Gasenzer}
\affiliation{Kirchhoff-Institut f{\"u}r Physik, Universit{\"a}t Heidelberg, Im Neuenheimer Feld 227, 69120 Heidelberg, Germany}
\affiliation{ExtreMe Matter Institute EMMI, GSI Helmholtzzentrum f{\"u}r Schwerionenforschung, 64291 Darmstadt, Germany}

\author{Matthew J Davis}
\affiliation{School of Mathematics and Physics, The University of Queensland, Brisbane QLD 4072, Australia}
\affiliation{JILA, University of Colorado, 440 UCB, Boulder, Colorado 80309, USA}

\begin{abstract}
We use the coordinate Bethe ansatz to exactly calculate matrix elements between eigenstates of the Lieb--Liniger model of one-dimensional bosons interacting via a two-body delta-potential. We investigate the static correlation functions of the zero-temperature ground state and their dependence on interaction strength, and analyze the effects of system size in the crossover from few-body to mesoscopic regimes for up to seven particles. We also obtain time-dependent nonequilibrium correlation functions for five particles following quenches of the interaction strength from two distinct initial states.  One quench is from the non-interacting ground state and the other from a correlated ground state near the strongly interacting Tonks-Girardeau regime. The final interaction strength and conserved energy are chosen to be the same for both quenches.  The integrability of the model highly constrains its dynamics, and we demonstrate that the time-averaged correlation functions following quenches from these two distinct initial conditions are both nonthermal and moreover distinct from one another.
\end{abstract}

%{\noindent{\it Keywords\/}: Lieb--Liniger model, Bethe ansatz, one-dimensional quantum gases, few-body systems}

\pacs{02.30.Ik, 67.85.-d, 05.30.Jp}

\date{\today}

\maketitle

\section{Introduction}
The Lieb--Liniger model of a one-dimensional Bose gas with repulsive delta-function interactions is a paradigmatic example of an exactly solvable continuous, integrable many-body quantum system~\cite{Lieb1963a,*Lieb1963b}. In particular, it has served as the context for the development of theoretical tools that have subsequently been widely applied in the study of integrable systems, such as the so-called ``thermodynamic Bethe ansatz'' functional representation, which provides the exact equation of state, excitation spectrum~\cite{Lieb1963a,Lieb1963b}, and bulk parameters~\cite{Yang1969} of the system in the thermodynamic limit.  However, the calculation of correlation functions from the exact solutions provided by the Bethe ansatz is notoriously difficult. 

At zero temperature, exact closed-form solutions for some equilibrium correlation functions are known in the Tonks--Girardeau limit of infinite interaction strength~\cite{Girardeau1960,Schultz1963,Lenard1964,Vaidya1979,Miwa1981}. This comparatively tractable limit also allows for some strong-coupling expansion results for large but finite interactions~\cite{Miwa1981,Creamer1986,Cherny2006,Forrester2006}. In the opposite weakly interacting quasi-condensate regime, a mean-field approach can be used to describe the system~\cite{Pitaevskii2003} and a Bogoliubov method can be used to determine the low-lying excitation spectrum~\cite{Mora2003}, relying on small density fluctuations.
Fewer results are available for intermediate interaction strengths, away from the strongly-interacting and weakly-interacting regimes.  The development of the Luttinger liquid description of quantum fluids~\cite{Haldane1981} and the related formalism of conformal field theory~\cite{Cardy1984,Berkovich1991} have lead to the prediction of power-law scaling for first-order correlations at large distances, with an exponent given in terms of the equation of state that is known exactly from the thermodynamic Bethe ansatz~\cite{Cazalilla2004}. 
The algebraic Bethe ansatz provides a determinantal representation of correlations, from which their asymptotic behavior can be extracted~\cite{Korepin1993}.
More recently, exact expressions for local second- and third-order correlations~\cite{Gangardt2003a,Gangardt2003b,Cheianov2006}, together with exact results for the one-body correlation function at asymptotically short distances~\cite{Olshanii2003} in terms of the equation of state have been derived.

Away from the asymptotic short- and long-range regimes, the behavior of correlation functions is less well known. For intermediate interaction strengths and arbitrary length scales one must resort to numerics to determine the correlation functions. Results for the latter have been obtained using numerical methodologies including quantum Monte Carlo~\cite{Astrakharchik2003,Astrakharchik2006}, and density matrix renormalization group approaches~\cite{Schmidt2007}. A recently developed, integrability-based approach combines the decomposition of correlation functions into sums over matrix elements (form factors) of certain simple operators between Bethe ansatz eigenstates~\cite{Caux2006,CauxABACUS}. This approach has generated results, for example, for static and dynamical equilibrium correlations at zero and finite temperature~for systems of up to $N \approx 100$ particles~\cite{Panfil2014}.
Other finite temperature results for correlation functions have been obtained using imaginary time stochastic gauge methods~\cite{Drummond2004,Deuar2009}, taking the non-relativistic limit of a relativistic field theory~\cite{Kormos2010}, utilizing Fermi--Bose mapping for the strongly interacting gas~\cite{Lenard1966,Cherny2006,Vignolo2013}, employing perturbative expansions in temperature and interaction strength~\cite{Sykes2008}, as well as combining the thermodynamic Bethe ansatz with the Hellmann--Feynman theorem~\cite{Kheruntsyan2003}.

Experiments with ultracold quantum gases are able to realize effectively one-dimensional systems by tightly confining the gas in two of the three spatial dimensions, either using optical lattice potentials or atom-chip traps~\cite{Kinoshita2005,Kinoshita2006,Hofferberth2007,Gring2012,Langen2013,Haller2009,Fabbri2014,Ott2012,Ott2013,Fabbri2014,Fabbri2009,Paredes2004,VanDruten2008,Armijo2010,Armijo2011,Jacqmin2011,Meinert2015}. These experiments are now probing the predictions of the Lieb--Liniger model. 
The configurability of quantum-gas experiments allows for so-called \emph{quenches} of the system, in which Hamiltonian parameters of the system are abruptly changed, and thus for the study of the Lieb-Linger model out of equilibrium, providing even greater challenges for theory.

The dynamically evolving correlations of the Lieb--Liniger gas in nonequilibrium scenarios are currently a topic of significant interest, and a number of theoretical approaches have been applied. Notable examples include exact diagonalization under a low momentum cutoff~\cite{Berman2004,Wintergerst2013a,Wintergerst2013b,Kanamoto2003,Kanamoto2006}, mapping of the hard-core Tonks--Girardeau gas to free spinless fermions~\cite{Rigol2004,Rigol2006,Rigol2007,Cassidy2011,Gramsch2012,He2013,Wright2013,Kormos2014}, phase-space methods~\cite{Deuar2006}, dynamic Bogoliubov-like approximations~\cite{Natu2013} and tensor-network methods~\cite{Muth2010a,Muth2010b}.
References~\cite{Gasenzer2005,Berges2007,Branschadel2008,Gasenzer2008} employed nonperturbative approximative functional-integral methods, while in Ref.~\cite{Cazalilla2009} a dynamical Luttinger-liquid approach was taken.
Other calculations make explicit use of the integrability of the system.  These are based on various Bethe ansatz approaches, and include utilizing Fermi-Bose mapping~\cite{Collura2013a,Collura2013b} and strong coupling expansions of the coordinate Bethe ansatz wave function~\cite{Buljan2008,Jukic2008,Pezer2009}, 
combining the algebraic Bethe ansatz with other numerical methods~\cite{Gritsev2010,Sato2012,Mossel2010},
and using the Yudson contour-integral representation for infinite-length systems~\cite{Iyer2012,Iyer2013}.
Recently, it was conjectured that the dynamics following an interaction strength quench are captured by a thermodynamic Bethe ansatz saddle point state and excitations around it --- the so-called quench action approach~\cite{Caux2013,DeNardis2014,Mussardo2013,DeNardis2014b,DeNardis2015,Piroli2015a}.
In the spirit of the methodology of Refs.~\cite{Caux2007,Caux2006}, Gritsev \emph{et al.}~\cite{Gritsev2010} investigated a quench from $\gamma=0 \rightarrow \infty$ by combining algebraic Bethe ansatz expressions for form factors with truncated sums over states, and employing Monte Carlo summation over the eigenstate components of the initial state.

In this paper we take a different approach, and calculate correlation functions of the Lieb-Linger model, both in and out of equilibrium, by calculating matrix elements between Lieb--Liniger eigenstates directly within the coordinate Bethe ansatz formalism.  Given the known  expressions for the coordinate-space forms of Lieb--Liniger eigenstates, we generate symbolic expressions for matrix elements of operators between these states in terms of the Bethe rapidities. The numerically obtained values of the rapidities can then be substituted to yield essentially numerically exact values for the matrix elements. 

In our previous work we applied this methodology to quenches from the ideal gas ground state to positive $\gamma$ for up to $N=5$ particles~\cite{Zill2015a}.  In Sec.~\ref{sec:Model}  we provide the details of the methodology, and describe how it can be used to calculate the matrix elements of the Lieb--Liniger eigenstates.  These symbolic expressions, and thus the computational cost of evaluating them, grow combinatorially with particle number, restricting the method to systems of only a few particles. However for small particle numbers $N \leq 7$ we obtain numerically exact results for ground-state correlations, which are described in Sec.~\ref{sec:GroundState}. Our results demonstrate that local correlations in the strongly interacting regime are already close to their thermodynamic-limit values for these few-body to mesoscopic systems.

An additional advantage of our methodology is that it can also calculate overlaps between Lieb--Liniger eigenstates corresponding to \emph{any} two interaction strengths, which allows us to study the dynamics of quenches of the interaction strength between arbitrary values. In Sec.~\ref{sec:Dynamics} we utilize this property to study the effects of integrability on the relaxation of the Lieb-Liniger model following such a quench.  In particular, we compare two nonequilibrium quench scenarios with the same final Hamiltonian and state energy, but beginning from starkly different initial states.  Statistical mechanics would predict that the system would relax to the same thermal state in both cases, but due to the integrability of the Lieb-Linger model not only are the time-averaged states following the two quenches non-thermal, they are also distinct.  After characterizing and comparing the nonequilibrium dynamics following both quenches, we conclude in Sec.~\ref{sec:conclusions}.

\section{Coordinate Bethe-ansatz methodology}\label{sec:Model}
\subsection{Lieb--Liniger model eigenstates}
The Lieb--Liniger model~\cite{Lieb1963a,Lieb1963b} describes a system of $N$ indistinguishable bosons subject to a delta-function interaction potential in a periodic one-dimensional (1D) geometry of length $L$. 
We work in units such that $\hbar=1$ and the particle mass $m=1/2$, and so the Hamiltonian of this system reads
\begin{equation}\label{eq:LLmodel}
    \hat{H} = - \sum_{i=1}^{N} \frac{\partial^2}{\partial x^2_i} + 2c \sum_{i<j}^{N} \delta(x_i - x_j),
\end{equation}
where $c$ is the interaction strength. 
The coordinate Bethe ansatz yields eigenstates $|\{\lambda_j\}\rangle$ of Hamiltonian~\eqref{eq:LLmodel} with spatial representation~\cite{Korepin1993}
\begin{align}\label{eq:eigenfunction_unordered}
    \zeta_{\{\lambda_j\}}(\{x_i\}) &\equiv \langle \{x_i\}|\{\lambda_j\}\rangle \nonumber \\
 &= \, A_{\{\lambda_j\}} \sum_{\sigma}  \; \mathrm{exp}\Big[ i \sum_{m=1}^{N}x_m \lambda_{\sigma(m)}\Big] \nonumber \\
 &\qquad\times \prod_{k>l} \left( 1 - \frac{ic \, \mathrm{sgn}(x_k - x_l) }{\lambda_{\sigma(k)}-\lambda_{\sigma(l)}} \right),
\end{align}
where the rapidities $\lambda_j$ (or quasimomenta) are solutions of the Bethe equations
\begin{equation}\label{eq:Betheeq}
    \lambda_j =\frac{2\pi}{L} m_j - \frac{2}{L} \sum_{k=1}^{N} \mathrm{arctan} \left( \frac{\lambda_j-\lambda_k}{c} \right).
\end{equation}
The quantum numbers $m_j$ are any $N$ distinct integers (half-integers) in the case that $N$ is odd (even)~\cite{Yang1969}, and $\sum_{\sigma}$ denotes a sum over all $N!$ permutations $\sigma=\{\sigma(j)\}$ of $\{1,2, \dots,N\}$. The normalization constant reads~\cite{Korepin1993}
\begin{equation}\label{eq:norm}
    A_{\{\lambda_j\}} = \frac{   \prod_{k>l}(\lambda_k-\lambda_l)}{ \big[ N! \;  \mathrm{det}\{M_{\{\lambda_j\}}\} \;\prod_{k>l} [(\lambda_k-\lambda_l)^2 + c^2 ] \big]^{1/2}} \; ,
\end{equation}
where $M_{\{\lambda_j\}}$ is the $N \times N$ matrix with elements
\begin{align}\label{eq:Gaudin_mtx}
    [M_{\{\lambda_j\}}]_{kl} &= \delta_{kl} \Big( L + \sum_{m=1}^{N} \frac{2c}{c^2 + (\lambda_k-\lambda_m)^2} \Big)\nonumber \\
&\qquad - \frac{2c}{c^2 + (\lambda_k-\lambda_l)^2}.
\end{align}
The rapidities determine the total momentum $P=\sum_{j=1}^{N} \lambda_j$ and energy $E = \sum_{j=1}^{N} \lambda_j^2$ of the system in each eigenstate. The ground state of the system corresponds to the set of $N$ rapidities that minimize $E$ and constitute the (pseudo-)Fermi sea of the 1D Bose gas~\cite{Korepin1993}. The Fermi momentum 
\begin{equation}\label{eq:kF}
    k_\mathrm{{F}} = \frac{2\pi}{L}\frac{N-1}{2}
\end{equation}
is the magnitude of the largest rapidity occurring in the ground state in the Tonks--Girardeau limit of strong interactions~\cite{Girardeau1960}.  
The only parameter of the Lieb--Liniger model in the thermodynamic limit is the dimensionless coupling $\gamma \equiv c/n$, where $n \equiv N/L$ is the 1D density. In finite systems, physical quantities also depend on the particle number $N$ (see, e.g., Sec.~\ref{subsec:FiniteSize}), whereas the length $L$ of our system, and therefore also the density $n$, are arbitrary. Consequently, in this article we will specify both $N$ and $\gamma$. Unless specified otherwise, we measure time in units of $k_\mathrm{{F}}^{-2}$, energy in units of $k_\mathrm{{F}}^2$, and length in units of $k_\mathrm{{F}}^{-1}$.

\subsection{Calculation of correlation functions and overlaps}\label{subsec:correlation_functions}
As the eigenstates $|\{\lambda_j\}\rangle$ form a complete basis~\cite{Dorlas1993} for the state space of the Lieb--Liniger model, the expectation value $\langle \hat{O} \rangle_{t}= \mathrm{Tr}\{\hat{\rho}(t) \hat{O}\}$ of an arbitrary operator $\hat{O}$ in a Schr{\"o}dinger-picture density matrix $\hat{\rho}(t)$ can be expressed as a sum of matrix elements of $\hat{O}$ between the states $|\{\lambda_j\}\rangle$.  In particular, in a pure state $|\psi(t)\rangle = \sum_{\{\lambda_j\}} C_{\{\lambda_j\}}(t) |\{\lambda_j\}\rangle$ we have
\begin{align}\label{eq:observable}
     \langle \hat{O} \rangle_{t} &\equiv \langle \psi(t) |\hat{O}| \psi(t) \rangle \nonumber \\
    &=\sum_{\{\lambda_j\}}\sum_{\{\lambda'_j\}} C^*_{\{\lambda'_j\}}(t) C^{}_{\{\lambda_j\}}(t)  \langle \{\lambda'_j\} | \hat{O} | \{\lambda_j\} \rangle,
\end{align}
whereas in a statistical ensemble with density matrix $\hat{\rho}_\mathrm{SE} = \sum_{\{\lambda_j\}} \rho^\mathrm{SE}_{\{\lambda_j\}}|\{\lambda_j\}\rangle\langle\{\lambda_j\}|$, we find
\begin{align}\label{eq:observable_mixed}
    \langle \hat{O} \rangle &= \sum_{\{\lambda_j\}} \rho^\mathrm{SE}_{\{\lambda_j\}} \langle \{\lambda_j\}|\hat{O}|\{\lambda_j\}\rangle.
\end{align}
In this article, we focus in particular on the normalized $m^\mathrm{th}$-order equal-time correlation functions 
\begin{align}
    &g^{(m)}(x_1, \dots, x_m, x_1', \dots, x_m'; t) \nonumber \\
    &\quad \equiv \frac{\left\langle \hat{\Psi}^\dagger (x_1) \cdots \hat{\Psi}^\dagger (x_m) \hat{\Psi}(x_1')\cdots \hat{\Psi}(x_m')\right\rangle}{\left[\langle \hat{n}(x_1)\rangle \cdots \langle \hat{n}(x_m)\rangle \langle \hat{n}(x_1')\rangle \cdots\langle \hat{n}(x_m')\rangle\right]^{1/2}},
\end{align}
where $\hat{\Psi}^{(\dagger)}(x)$ is the annihilation (creation) operator for the Bose field and $\hat{n}(x) \equiv \hat{\Psi}^\dagger(x) \hat{\Psi}(x)$. 
Here and in the following we drop the time index $t$ of the state vectors.

Since the Hamiltonian we consider in this article is translationally invariant along the periodic volume of length $L$, the mean density $\langle \hat{n}(x) \rangle \equiv n$ is constant in both time and space, and $g^{(m)}(x_1, \dots, x_m, x_1', \dots, x_m'; t) = \langle \hat{\Psi}^\dagger (x_1) \cdots \hat{\Psi}^\dagger (x_m) \hat{\Psi}(x_1')\cdots \hat{\Psi}(x_m')\rangle /n^m$. 
The correlation functions $g^{(m)}(x_1, \dots, x_m, x_1', \dots, x_m'; t)$ can therefore be expressed as the expectation values of the operators $\hat{g}^{(m)}(x_1, \dots, x_m, x_1', \dots, x_m') \equiv \hat{\Psi}^\dagger (x_1) \cdots \hat{\Psi}^\dagger (x_m) \hat{\Psi}(x_1')\cdots \hat{\Psi}(x_m')/n^m$.  
We note that for the same reasons as above the matrix elements $\langle \{\lambda'_j\} | \hat{g}^{(m)}(x_1,\dots,x_m,x_1',\dots,x_m')| \{\lambda_j\}\rangle$ are invariant under global coordinate shifts $x\to x + d$ and thus, without loss of generality, we can set one of the spatial variables to zero. 
For the first-order correlation function, the matrix elements are 
\begin{align}\label{eq:G1mtx_elt_fundamental}
   & \langle \{\lambda'_j\}| \hat{g}^{(1)}(0,x) | \{\lambda_j\} \rangle \equiv \langle \{\lambda'_j\}| \hat{\Psi}^\dagger(0) \hat{\Psi}(x) | \{\lambda_j\} \rangle \nonumber \\
    &\qquad = \frac{N}{n} \int dx_1\cdots dx_{N-1}\, \zeta^*_{\{\lambda'_j\}}(0,x_1,\dots,x_{N-1})\, \nonumber \\
             &\qquad \quad \quad  \times \zeta_{\{\lambda_j\}}(x,x_1,\dots,x_{N-1}).
\end{align}
The evaluation of the integral in Eq.~\eqref{eq:G1mtx_elt_fundamental} is complicated by the sign function in Eq.~\eqref{eq:eigenfunction_unordered} and the associated nonanalyticities in $\zeta_{\{\lambda_j\}}(\{x_i\})$ where any two particle coordinates $x_k$ and $x_l$ coincide.
However, we can use the Bose symmetry of the wave function $\zeta_{\{\lambda_j\}}(\{x_i\})$ to reexpress this matrix element as a sum of integrals
\begin{widetext}
\begin{align}
    \langle \{\lambda'_j\}| \hat{g}^{(1)}(0,x) | \{\lambda_j\} \rangle &= \frac{N!}{n} \sum_{\ell=0}^{N-1} \int_{\mathcal{R}_{N-1,\ell}(x)} dx_1\cdots dx_{N-1}\, \zeta^*_{\{\lambda'_j\}}(0,x_1,\dots,x_{N-1}) \, \zeta_{\{\lambda_j\}}  (x_1,\dots,x_\ell,x,x_{\ell+1},\dots,x_{N-1}),
\end{align}
over the ordered domains~\cite{Forrester2006} 
\begin{align}  
    \mathcal{R}_{M,j}(x):\quad 0\leq x_1 < \cdots < x_j < x < x_{j+1} < \cdots < x_M \leq L.
\end{align}
Substituting the coordinate-space form [Eq.~\eqref{eq:eigenfunction_unordered}] of the Lieb-Liniger eigenfunctions, we obtain
\begin{align}\label{eq:G1mtxelt}
    \langle \{\lambda'_j\}| \hat{g}^{(1)}&(0,x) | \{\lambda_j\} \rangle = \frac{N!}{n}  A^{}_{\{\lambda^{}_j\}} A^*_{\{\lambda'_j\}} \sum_{\sigma} \sum_{\sigma'} \prod_{j>k} \Big(1 - \frac{ic}{\lambda_{\sigma(j)}-\lambda_{\sigma(k)}}\Big) \prod_{j'>k'} \Big(1 + \frac{ic}{\lambda'_{\sigma'(j')}-\lambda'_{\sigma'(k')}}\Big) \nonumber \\
    &\qquad \times \sum_{\ell=0}^{N-1} \exp(i\lambda_{\sigma(\ell+1)} x) \int_{\mathcal{R}_{N-1,\ell}(x)} dx_1\cdots dx_{N-1} \exp\left(i\sum_{m=1}^{N-1} ( \lambda_{\sigma^{(\ell+1)}(m)} - \lambda'_{\sigma'(m+1)})x_m\right),
\end{align}
where $\sigma^{(\ell+1)} = (\sigma(1),\dots,\sigma(\ell),\sigma(\ell+2),\dots, \sigma(N))$.
The matrix elements of the second-order correlation operator $\hat{g}^{(2)}(0,x)\equiv \hat{\Psi}^\dagger(0)\hat{\Psi}^\dagger(x)\hat{\Psi}(x)\hat{\Psi}(0)/n^2$ are similarly given by
\begin{align}\label{eq:G2mtxelt} 
    \langle \{\lambda'_j\}|& \hat{g}^{(2)}(0,x)| \{\lambda_j\} \rangle = \frac{N!}{n^2} A^{}_{\{\lambda^{}_j\}} A^*_{\{\lambda'_j\}} \sum_{\sigma} \sum_{\sigma'} 
\prod_{j>k} \Big(1 - \frac{ic}{\lambda_{\sigma(j)}-\lambda_{\sigma(k)}}\Big) \prod_{j'>k'} \Big(1 + \frac{ic}{\lambda'_{\sigma'(j')}-\lambda'_{\sigma'(k')}}\Big)  \nonumber \\
& \times \sum_{\ell=0}^{N-2} \exp\left(i(\lambda_{\sigma(\ell+2)} - \lambda'_{\sigma'(\ell+2)})x\right) \int_{\mathcal{R}_{N-2,\ell}(x)} dx_1\cdots dx_{N-2} \exp\left(i \sum_{m=1}^{N-2} (\lambda_{\sigma^{(1,\ell+2)}(m)} - \lambda'_{\sigma'^{(1,\ell+2)}(m)}) x_m\right),
\end{align}
where $\sigma^{(1,\ell+2)} = (\sigma(2),\dots,\sigma(\ell+1),\sigma(\ell+3),\dots,\sigma(N))$ and $\sigma'^{(1,\ell+2)}$ is defined analogously in terms of the elements of $\sigma'$.  In the limit $x\to 0$ this expression simplifies somewhat, and in general the matrix elements of the local $m^\mathrm{th}$-order correlation operator $\hat{g}^{(m)}(0) \equiv [\hat{\Psi}^\dagger(0)]^m [\hat{\Psi}(0)]^m/n^m$ are given by the expression
\begin{align}\label{eq:Gm0mtxelt} 
    \langle \{\lambda'_j\}| \hat{g}^{(m)}(0) | \{\lambda_j\} \rangle &= \frac{N!}{n^m}  A^{}_{\{\lambda^{}_j\}} A^*_{\{\lambda'_j\}} \sum_{\sigma} \sum_{\sigma'} \prod_{j>k} \Big(1 - \frac{ic}{\lambda_{\sigma(j)}-\lambda_{\sigma(k)}}\Big) \prod_{j'>k'} \Big(1 + \frac{ic}{\lambda'_{\sigma'(j')}-\lambda'_{\sigma'(k')}}\Big) \nonumber \\
&\qquad \times \sum_{\ell=0}^{N-m} \int_{\mathcal{R}_{N-m}} dx_1\cdots dx_{N-m} \exp\left(i \sum_{n=1}^{N-m} (\lambda_{\sigma(m+n)} - \lambda'_{\sigma'(m+n)}) x_n\right),
\end{align}
where the domain $\mathcal{R}_M : 0 \leq x_1 < x_2 < \cdots < x_M \leq L$.  We note, moreover, that Eqs.~\eqref{eq:G1mtxelt}--\eqref{eq:Gm0mtxelt} include as degenerate cases the diagonal matrix elements (cf.\ Ref.~\cite{Forrester2006}) appropriate to the calculation of correlations in the ground state (Sec.~\ref{sec:GroundState}) and in statistical ensembles (Sec.~\ref{sec:Dynamics}).

The calculation of correlation functions from Eqs.~\eqref{eq:G1mtxelt}--\eqref{eq:Gm0mtxelt} involves the evaluation of integrals of the general form
\begin{align}\label{eq:generic_integral}
    \int_{\mathcal{R}_{M,\ell}(x)} \! dx_1\cdots dx_M \exp\left(i\sum_{m=1}^M \kappa_m x_m\right) &= \int_x^L\!dx_M e^{i\kappa_M x_M} \int_x^{x_M}\!dx_{M-1} e^{i\kappa_{M-1} x_{M-1}} \cdots \int_x^{x_{\ell+2}}\!\!dx_{\ell+1} e^{i\kappa_{\ell+1} x_{\ell+1}}  \nonumber \\
&\qquad\times \int_0^x\!dx_\ell e^{i\kappa_\ell x_\ell} \int_0^{x_\ell}\!dx_{\ell-1} e^{i\kappa_{\ell-1} x_{\ell-1}} \cdots \int_0^{x_2}\!dx_1 e^{i\kappa_1 x_1},
\end{align}
\end{widetext}
where (for the repulsive interactions $c>0$ considered in this article) the $\kappa_m$ are real numbers.  A single closed form for this integral does not exist, as in general one or more $\kappa_m$ may vanish, and this must be handled separately from the case of $\kappa_m\neq 0$.  However, given knowledge of the \emph{particular} sets of rapidities $\{\lambda_j\}$ and $\{\lambda'_j\}$ (and permutations $\sigma$ and $\sigma'$), and thus of the locations of zero exponents $\kappa_m=0$ in Eq.~\eqref{eq:generic_integral}, 
each individual integral of this form can be reduced to an algebraic expression in terms of $\{\kappa_m\}$.  
More specifically, each successive integration $\int dx_m$ yields a term (involving, in general, $x_{m+1}$) arising from the primitive integral~\cite{Arfken2005}
\begin{align}
\int dx \; x^p e^{ikx} &=  -(i/k)^{p+1}\Gamma(p+1,-ikx)\nonumber \\
    &= - p!(i/k)^{p+1} e^{ikx} \sum_{s=0}^p \frac{(-ikx)^s}{s!}  ,
\end{align}
in the case that $\kappa_m$ is nonzero, or from $\int dx \, x^p$ otherwise.  In our calculations, the construction of algebraic expressions for the integrals occurring in Eqs.~\eqref{eq:G1mtxelt}--\eqref{eq:Gm0mtxelt} in terms of the rapidities $\lambda_j$ is efficiently performed by a simple computer algorithm that accounts for and combines the symbolic terms that arise from these successive reductions. We note that, e.g., each matrix element $\langle \{\lambda'_j\}| \hat{g}^{(1)}(0,x) | \{\lambda_j\} \rangle$ is a sum of $N$ integrals over $(N-1)$-dimensional domains and that the integrand in each case comprises $(N!)^2$ terms~\cite{Forrester2006}, illustrating the dramatically increasing computational cost of evaluating correlation functions with increasing $N$. Nevertheless, the explicit closed-form expression for the integral produced by our algorithm can be evaluated to obtain a numerically exact result by substituting in the values of the rapidities. The latter are obtained by solving Eq.~\eqref{eq:Betheeq} numerically using Newton's method, starting in the Tonks--Girardeau regime of strong interactions $\gamma \gg1$ and iteratively progressing to smaller values of $\gamma$ using initial guesses given by linear extrapolation of the solutions at stronger interaction strengths.

We note that this algorithmic approach also provides for the efficient and accurate calculation of the overlaps $\langle \{\lambda_j\} |\{\mu_j\}\rangle$ between eigenstates of Hamiltonian~\eqref{eq:LLmodel} corresponding to different values of $\gamma$, which we make use of in our analysis of nonequilibrium dynamics in Sec.~\ref{sec:Dynamics}. In particular, the overlap between an arbitrary eigenstate $|\{\lambda_j\} \rangle$ of $\hat{H}$ at a finite interaction strength $\gamma>0$ and the noninteracting ground state $|0\rangle$, with constant spatial representation $\langle \{x_i\}|0\rangle = L^{-N/2}$, is simply given by
\begin{widetext}
\begin{align}\label{eq:overlaps}
    \langle \{\lambda_j\} |0\rangle &= \frac{N!}{L^{N/2}} A_{\{\lambda_j\}} \sum_{\sigma} \prod_{j>k} \Big(1 + \frac{ic}{\lambda_{\sigma(j)}-\lambda_{\sigma(k)}}\Big) \int_{\mathcal{R}_N} dx_1\cdots dx_N \exp\left(-i\sum_{n=1}^N \lambda_{\sigma(n)}x_n\right),
\end{align}
\end{widetext}
which can easily be evaluated semi-analytically using our algorithm. In practice we find that the results we obtain for the overlaps from our evaluation of Eq.~\eqref{eq:overlaps} agree with the recently derived closed-form expressions for these quantities~\cite{DeNardis2014,Brockmann2014a,Brockmann2014b,Brockmann2014c}, which imply in particular that $\langle \{\lambda_j\} |0\rangle \propto 1/\lambda_j^2$ as any $\lambda_j\to\infty$.

\section{Ground-State Correlation Functions}\label{sec:GroundState} 
As a first application of our methodology we calculate the correlation functions of the Lieb--Liniger model in the ground state for up to $N=7$ particles.  In this case, we need to evaluate only the diagonal elements of Eqs.~\eqref{eq:G1mtxelt}--\eqref{eq:Gm0mtxelt} in the ground-state wave function, thereby obtaining exact algebraic expressions for correlation functions in terms of the ground-state rapidities, which are themselves determined to machine precision (Sec.~\ref{subsec:correlation_functions}). 
The ground-state correlations of the Lieb--Liniger model have been considered extensively in previous works (see Refs.~\cite{Yurovsky2008,Cazalilla2011} and references therein), and we compare our exact mesoscopic results to those obtained with various other methods and approximations, for finite system sizes as well as in the thermodynamic limit. This allows us to clarify the utility and limitations of calculations, such as ours here and in Ref.~\cite{Zill2015a}, that involve only small particle numbers.

\subsection{First-order correlations}\label{subsec:FirstOrder} 
\begin{figure}
   \includegraphics[width=0.48\textwidth]{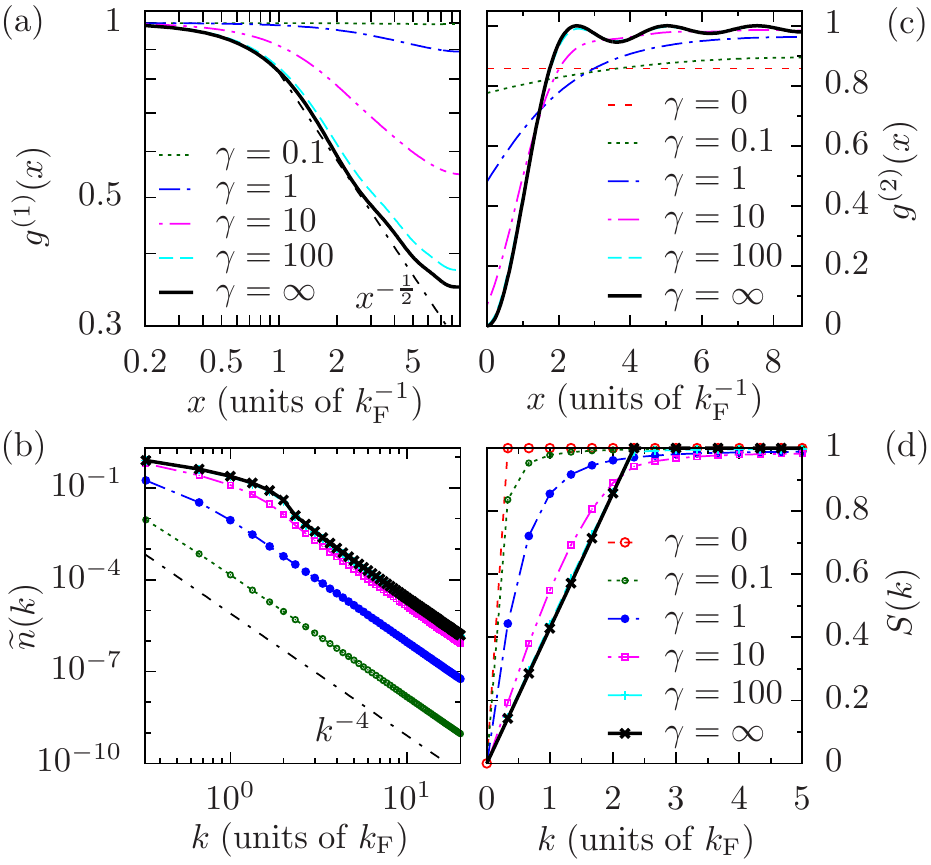}
    \caption{\label{fig:g1momdist} (Color online)  One- and two-body correlations in the Lieb--Liniger ground state, for $N=7$ particles. (a) Non-local first-order coherence $g^{(1)}(x)$. The black dot-dashed line indicates the asymptotic long-range behavior $g^{(1)}(x) \propto |x|^{-1/2}$ of a Tonks--Girardeau gas in the thermodynamic limit. (b) Corresponding zero-temperature momentum distribution $\widetilde{n}(k_j)$. The black dot-dashed line indicates the universal high-momentum power-law scaling $\widetilde{n}(k) \propto k^{-4}$ common to all positive interaction strengths~\cite{Olshanii2003}. (c) Non-local second-order coherence $g^{(2)}(x)$. (d) Corresponding static structure factor $S(k)$.}  
\end{figure}

We begin by considering the first-order correlation function $g^{(1)}(x)\equiv g^{(1)}(0,x)$ in the ground state of the Lieb--Liniger model.  In Fig.~\ref{fig:g1momdist}(a) we plot $g^{(1)}(x)$ for $N=7$ particles for a range of interaction strengths $\gamma$, which exhibits the expected decrease in spatial phase coherence with increasing $\gamma$~\cite{Cazalilla2004}.
As is well known, true long-range order, $\lim_{x\to\infty} g^{(1)}(x) = n_0 > 0$~\cite{Penrose1956,Yang1962}, is prohibited in an interacting homogeneous 1D Bose gas in the thermodynamic limit, even at zero temperature (see Ref.~\cite{Cazalilla2011} and references therein).  Indeed the Lieb--Liniger system is quantum critical at zero temperature, and the asymptotic long-range behavior of $g^{(1)}(x)$ is a power-law decay (so-called \emph{quasi}-long-range order)~\cite{Korepin1993}.

This power-law scaling of $g^{(1)}(x)$ is only expected to be realized at separations $x$ large compared to the healing length $\xi=1/\sqrt{\gamma}$ and, in a finite periodic geometry such as we consider here, is curtailed by the finite extent $L$ of the system~(see, e.g., Ref.~\cite{Cazalilla2004}). Indeed, for $\gamma=0.1$, the power-law decay is not visible in our finite-sized calculation, although as the interaction strength $\gamma$ increases $g^{(1)}(x)$ exhibits behavior consistent with power-law decay over an increasingly large range of $x$, see Fig.~\ref{fig:g1momdist}(a).  In particular, for $\gamma\gtrsim 10$, our results for $g^{(1)}(x)$ seem to converge toward the asymptotic scaling of the Tonks--Girardeau limit (black dot-dashed line) with increasing $\gamma$.   

Due to the translational invariance of our system, the first-order correlations of the Lieb--Liniger ground state are encoded in the momentum distribution
\begin{equation}\label{eq:mom_dist}
    \widetilde{n}(k_j) = n \int_{0}^{L} dx \; e^{-i k_j x} g^{(1)}(x),
\end{equation}
which, in our finite periodic geometry, is only defined for discrete momenta $k_j=2\pi j / L$, with $j$ an integer.  In Fig.~\ref{fig:g1momdist}(b) we plot the momentum distributions $\widetilde{n}(k_j)$ corresponding to the first-order correlation functions $g^{(1)}(x)$ shown in Fig.~\ref{fig:g1momdist}(a). 
The first feature that we note in Fig.~\ref{fig:g1momdist}(b) is that for all interaction strengths, $\widetilde{n}(k)$ exhibits a power-law decay $\widetilde{n}(k)\propto k^{-4}$ (dot-dashed black line) at high momenta. 
This is a universal result for delta-function interactions in one dimension~\cite{Olshanii2003,Caux2007,Barth2011} (and indeed also in higher dimensions~\cite{Tan2008}).  The effects of the finite extent $L$ of the system on the first-order correlations are again evident in this momentum-space representation. For $\gamma=0.1$, no deviation from the $\propto k^{-4}$ scaling is observed for the smallest (nonzero) momenta $k_j$ that can be resolved in the periodic geometry. For larger values of the interaction strength, $\widetilde{n}(k)$ departs from the $\propto k^{-4}$ scaling at increasingly large values of $k$ with increasing $\gamma$, and develops a hump at momenta near $k_\mathrm{{F}}$ for $\gamma \protect\gtrsim 10$~\cite{Caux2007}.  
We note that although the small-$k$ behavior of $\protect\widetilde{n}(k)$ tends towards the $\propto k^{-1/2}$ scaling exhibited by the Tonks--Girardeau gas in the thermodynamic limit, the rounding off of the power-law decay of $g^{(1)}(x)$ as $x\protect\to L/2$ precludes $\protect\widetilde{n}(k)$ from reaching the known asymptotic $k\protect\to 0$ behavior in our finite geometry.

\subsection{Second-, third-, and fourth-order correlations}\label{subsec:SecondOrder} 
In Fig.~\ref{fig:g1momdist}(c), we present the nonlocal second-order coherence $g^{(2)}(x)\equiv g^{(2)}(0,x,x,0)$, which provides a measure of density-density correlations, for $N=7$ particles at a range of interaction strengths $\gamma$.  In the limiting case of an ideal gas ($\gamma=0$), the ground state of the system is a Fock state of $N$ particles in the zero-momentum single-particle mode, and the second-order coherence $g^{(2)}_{\gamma=0}(x) = 1-N^{-1}$ (horizontal dashed line) is therefore independent of $x$. 
As the interaction strength $\gamma$ is increased, the second-order coherence is increasingly suppressed at zero spatial separation and correspondingly enhanced at separations $x \gtrsim 2 k_\mathrm{{F}}^{-1}$.  Oscillations in $g^{(2)}(x)$ develop at finite $x$ as the system enters the strongly interacting regime $\gamma\gg 1$~\cite{Korepin1993,Cherny2006} and, in particular, for $\gamma=100$ (dashed cyan line), our numerical results are practically indistinguishable from the exact Tonks--Girardeau limit result (solid black line)~\cite{Girardeau1960}. 

An alternative representation of the second-order correlations of the ground state is given by the static structure factor $S(k)$, which is related to $g^{(2)}(x)$ by~\cite{Pitaevskii2003}
\begin{equation}\label{eq:structfact}
    S(k_j) = 1 + n \int_{0}^{L} dx \; e^{-i k_j x} \left[g^{(2)}(x) - 1\right].
\end{equation}
In Fig.~\ref{fig:g1momdist}(d) we present the structure factors $S(k)$ corresponding to the correlation functions $g^{(2)}(x)$ shown in Fig.~\ref{fig:g1momdist}(c).  For all values of $\gamma$, $S(0) = 0$ due to particle-number conservation and translational invariance.  In the ideal-gas limit (red circles) $S(k_j)=1$ for all nonzero $k_j$. In the opposite limit of a Tonks--Girardeau gas
\begin{equation}\label{eq:Sk_TG_limit}
    S_{\gamma=\infty}(k_j) = \left\{
    \begin{array}{cl}
        \frac{|k_j|(1-N^{-1})}{2k_\mathrm{{F}}} & \quad\quad |k_j| \leq 2k_\mathrm{{F}} \\
        1 & \quad\quad |k_j| > 2k_\mathrm{{F}},  
    \end{array} \right.  % \}
\end{equation}
which tends, in the thermodynamic limit, to the well-known result (see, e.g., Ref.~\cite{Cherny2006}) $S(k) = |k|/2 k_\mathrm{{F}}$ for $|k|\leq 2 k_\mathrm{{F}}$, and $S(k)=1$ for $|k| > 2 k_\mathrm{{F}}$.  Just as for $g^{(2)}(x)$, we observe that for $\gamma=100$ (cyan plus symbols), our numerical results for $S(k)$ are almost identical to the known exact expression [Eq.~\eqref{eq:Sk_TG_limit}] for the Tonks--Girardeau limit (black crosses).  For smaller values of $\gamma$ our mesoscopic results for $S(k)$ appear consistent with those of Refs.~\cite{Astrakharchik2003,Caux2006}, obtained using quantum Monte Carlo and algebraic-Bethe ansatz techniques, respectively.

%%%
\begin{figure}
   \includegraphics[width=0.48\textwidth]{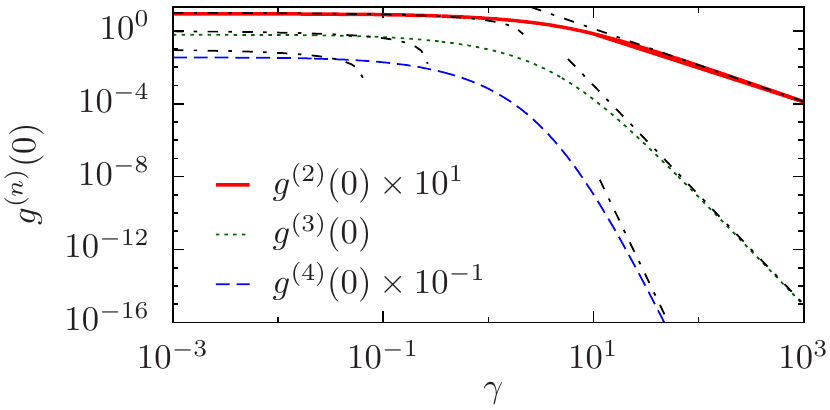}
    \caption{\label{fig:g23localgs} (Color online)  Interaction-strength dependence of the local second-, third- and fourth-order coherence in the Lieb--Liniger ground state, for $N=7$ particles. To aid visibility, we plot $g^{(2)}(0)$ scaled by a factor of $10^{1}$, and $g^{(4)}(0)$ scaled by a factor of $10^{-1}$.  Dot-dashed lines indicate asymptotic weak- ($\gamma \ll 1$) and strong-coupling ($\gamma \gg 1$) expressions for $g^{(2)}(0)$, $g^{(3)}(0)$ and $g^{(4)}(0)$ in the thermodynamic limit (see text).}
\end{figure}
%%%
We now focus in more detail on local correlation functions. We note that the local second-order coherence has recently been proposed as a measure of quantum criticality in the 1D boson system~\cite{Wang2013}, while the local third-order correlations have received increasing attention both in theory~\cite{Kormos2011} and experiment~\cite{Tolra2004,Armijo2010,Haller2011,Manning2013}. The local fourth-order correlations for the Lieb--Liniger model have also been investigated~\cite{Pozsgay2011b}.
In Fig.~\ref{fig:g23localgs}, we plot the local second-order coherence $g^{(2)}(0)$ (solid red line), together with the local third-order coherence $g^{(3)}(0)=\langle [ \hat{\Psi}^\dagger(0) ]^3 [\hat{\Psi}(0) ]^3 \rangle/n^3$ (dotted green line), and the local fourth-order coherence $g^{(4)}(0)=\langle [ \hat{\Psi}^\dagger(0) ]^4 [\hat{\Psi}(0) ]^4 \rangle/n^4$ (dashed blue line) for $N=7$ particles and a broad range of interaction strengths $\gamma$.
For comparison, we also plot the asymptotic results obtained in the Bogoliubov limit of weak interactions ($\gamma\to 0$) in the thermodynamic limit~\cite{Gangardt2003a,Mora2003} (left-hand dot-dashed lines). 
The numerical results for small $\gamma$ are broadly comparable to these thermodynamic-limit results.  However, for the small particle numbers considered here, the suppression of $g^{(2)}(0)$, $g^{(3)}(0)$, and $g^{(4)}(0)$ due to interactions in the limit of small $\gamma$ is overshadowed by the suppression due to the  finite population of the system~\cite{Cheianov2006}. 
At larger $\gamma$, the effects of interactions dominate, and the numerical results converge closely to the appropriate strong-coupling expressions~\cite{Gangardt2003a} (right-hand dot-dashed lines). We note, therefore, that the local correlations of the Lieb--Liniger ground state, and particularly their scaling with $\gamma$, appear to be quite insensitive to the infrared cutoff imposed by the finite extent of our system in the strongly interacting regime $\gamma \gg 1$.

\subsection{System-size dependence}\label{subsec:FiniteSize}  
The results we have obtained so far indicate that, as expected, the small size of our system leads to corrections to correlation functions as compared to their known asymptotic forms in the thermodynamic limit.  However, our results also suggest that the effects of finite system size are comparatively less important for local correlations, particularly in the limit of large interaction strengths $\gamma \gg 1$.  To further elucidate the potential significance of finite-size effects in our calculations of nonequilibrium dynamics~\cite{Zill2015a}, here we give a brief characterization of the dependence of correlation functions of the Lieb--Liniger ground state on the particle number $N$ at a fixed value of the interaction strength $\gamma$. 

Specifically we consider the case for $\gamma=10$, as this value places the system in the strongly interacting regime $\gamma \gg 1$ (which appears less sensitive to finite-size effects than the weakly interacting regime $\gamma \lesssim 1$), while still exhibiting significant deviations from the Tonks--Girardeau limit (see, e.g., Ref.~\cite{Cherny2006}).  Whereas elsewhere in this paper we quote momenta (lengths) in units of $k_\mathrm{{F}}$ ($k_\mathrm{{F}}^{-1}$), in comparing results between systems with different particle numbers $N$ we quote momenta (lengths) in units of $\pi n$ [$(\pi n)^{-1}$], so as to avoid a potentially misleading dependence of the unit of length on $N$ [cf. Eq.~\eqref{eq:kF}].
%%%
\begin{figure}
    \includegraphics[width=0.48\textwidth]{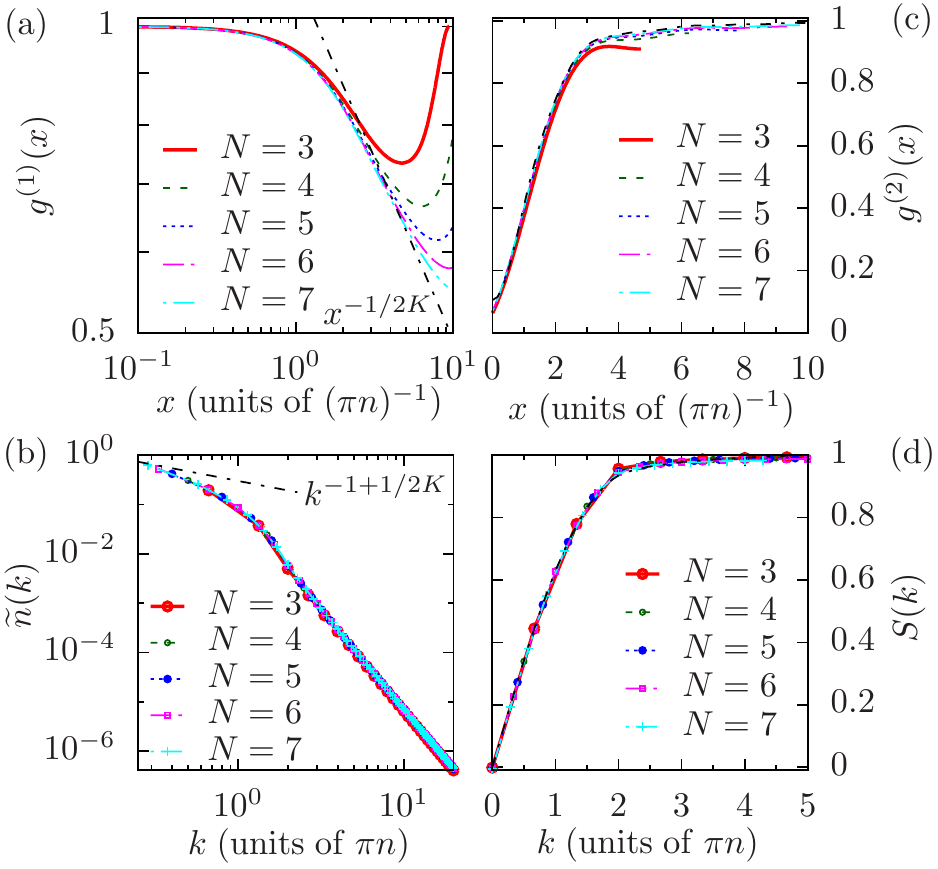}
    \caption{\label{fig:finitesizegs} (Color online)  Dependence of first- and second-order correlations in the Lieb--Liniger ground state on particle number $N$ for $\gamma=10$. (a) First-order correlation function $g^{(1)}(x)$. (b) Corresponding momentum distribution function $\widetilde{n}(k_j)$. Black dot-dashed lines in (a)~and~(b) indicate the asymptotic infrared scaling of $g^{(1)}(x)$ and $\widetilde{n}(k)$, respectively, with Luttinger parameter $K=1.40$ (see text). (c) Second-order correlation function $g^{(2)}(x)$. (d) Corresponding static structure factor $S(k)$. The black dot-dashed lines in (c) and (d) represent the phenomenological expressions of Ref.~\cite{Cherny2009} for $g^{(2)}(x)$ and $S(k)$ in the thermodynamic limit, respectively.}
\end{figure}
%%%

In Fig.~\ref{fig:finitesizegs}(a) we plot $g^{(1)}(x)$ for particle numbers $N=3,4,5,6,$ and $7$.  For small $x$, the curves fall nearly perfectly on one line. The same behavior can be observed for the large-$k$ tail of the corresponding momentum distribution $\widetilde{n}(k)$, which we plot in Fig.~\ref{fig:finitesizegs}(b). Indeed, at larger momenta $k \gtrsim 2 \pi n$, $\widetilde{n}(k)$ appears to exhibit a rapid collapse to a single curve with increasing $N$~\cite{Olshanii2003,Xu2015}. However, the differences in $\widetilde{n}(k)$ are so small that they can not be seen in Fig.~\ref{fig:finitesizegs}(b). 
For small momenta, our choice of units implies an increasing resolution with increasing particle number, specifically $k_1 = 2\pi/L \times (\pi n)^{-1} = 2/N$. However, this lowest resolvable momentum seems to fall on one line for increasing particle number, indicating that the infrared behavior of large systems can be at least partly accessed by our mesoscopic system sizes.

Luttinger-liquid theory predicts a long-range power-law decay $g^{(1)}(x)\propto |x|^{-1/2K}$, where the Luttinger parameter $K$ can be calculated from the thermodynamic limit of the Bethe ansatz solution (see, e.g., Refs.~\cite{Korepin1993,Cazalilla2004} and references therein).  For our parameters we have $K=1.40$, implying an asymptotic scaling $g^{(1)}(x) \propto |x|^{-0.357}$ [black dot-dashed line in Fig.~\ref{fig:finitesizegs}(a)]. This corresponds to a power-law behavior $\widetilde{n}(k)\propto |k|^{-1+1/2K} = |k|^{-0.643}$~\cite{Cazalilla2004} [dot-dashed line in Fig.~\ref{fig:finitesizegs}(b)] for small momenta. 
We note that this infrared scaling is a true many-body effect and as such does not show up for $N=2$ particles. Indeed, one can show analytically that, for $N=2$, the momentum distribution $\widetilde{n}(k) \propto (\lambda_1^2 - k^2)^{-2}$ and thus $k^{-4}$ is the highest power in the series expansion of $\widetilde{n}(k)$.

In Fig.~\ref{fig:finitesizegs}(c) we plot the nonlocal second-order coherence $g^{(2)}(x)$ for $\gamma=10$ and $N=3,4,5,6,$ and $7$.  The corresponding static structure factor $S(k)$ is shown in Fig.~\ref{fig:finitesizegs}(d).  In Fig.~\ref{fig:finitesizegs}(d) we also plot (black dot-dashed line) the form of $S(k)$ resulting from the phenomenological expression proposed in Ref.~\cite{Cherny2009} (see also Ref.~\cite{Golovach2009}).  This expression involves the limiting dispersions and edge exponents of the Lieb--Liniger model, which we obtain by numerically solving the appropriate integral equations~\cite{Lieb1963b, Imambekov2008}.  We also plot the corresponding prediction for $g^{(2)}(x)$ (black dot-dashed line) in Fig.~\ref{fig:finitesizegs}(c).  
We note that the numerical results for our mesoscopic systems are, in general, rather close to the phenomenological thermodynamic-limit expressions even for the relatively small particle numbers considered here.

\section{Application to nonequilibrium dynamics}\label{sec:Dynamics}
We now apply our methodology to the nonequilibrium dynamics of the Lieb--Liniger model. Specifically, we consider the evolution of a system, initially prepared in the ground state of Hamiltonian~\eqref{eq:LLmodel} with interaction strength $\gamma_0$, following an abrupt change, at time $t=0$, of the interaction strength to a distinct value $\gamma\neq\gamma_0$ --- a so-called ``interaction quench''. The evolution of the system following such a quench is generated by Hamiltonian~\eqref{eq:LLmodel} with interaction strength $\gamma$, which we denote by $\hat{H}(\gamma)$ hereafter.  
The time-evolving state is given at all times $t>0$ by
\begin{equation}\label{eq:psievolution}
|\psi(t)\rangle = \sum_{\{ \lambda_j\} } \, C_{\{ \lambda_j\} } \mathrm{e}^{-i E_{\{ \lambda_j\} }t} |\{ \lambda_j\} \rangle,
\end{equation}
where $|\{ \lambda_j\} \rangle$ are the eigenstates of $\hat{H}(\gamma)$ with energies $E_{\{ \lambda_j\} }$, and $C_{\{ \lambda_j\} } \equiv \langle \{ \lambda_j\}  | \psi_0 \rangle$ are the overlaps of the $|\{ \lambda_j\} \rangle$ with the initial state $| \psi_0 \rangle$.
The expectation value of an arbitrary operator $\hat{O}$ in the state $|\psi(t)\rangle$ is given by  
\begin{align} \label{eq:operatorevolution}
&\langle \hat{O} \rangle_{t} \equiv \langle \psi(t)| \hat{O} | \psi(t)\rangle \nonumber \\
&= \sum_{\{\lambda_j\}} \sum_{\{\lambda_j'\}} C_{\{\lambda_j'\}}^* C^{}_{\{\lambda_j\}} \, \mathrm{e}^{i (E_{\{ \lambda_j'\} } - E^{}_{\{ \lambda_j\} }) t} \, \langle \{\lambda_j'\} | \hat{O} | \{\lambda_j\} \rangle .
\end{align}
We use the methodology described in Sec.~\ref{sec:Model} to evaluate both the overlaps $C^{}_{\{\lambda_j\}}$ and the matrix elements $\langle \{\lambda_j'\} | \hat{O} | \{\lambda_j\} \rangle$ that appear in Eq.~\eqref{eq:operatorevolution}.

One of the features of our methodology is that it allows us to describe quenches between arbitrary interaction strengths.
In this paper we consider two interaction-strength quenches, from different initial interaction strengths $\gamma_0$, to a common final value of the coupling $\gamma$.  Specifically, we consider a quench from the non-interacting limit $\gamma_0=0$ (similar to those previously studied in Refs.~\cite{Gritsev2010,DeNardis2014,Zill2015a,Kormos2013,Sotiriadis2014,Calabrese2014,Kormos2014,Collura2014}) and a quench from the correlated ground state obtained for a strong interaction strength $\gamma_0=100$.  As $\hat{H}(\gamma)$ is time independent following the quench, energy is conserved during the dynamics.  We choose the final interaction strength after the two quenches such that the postquench energy is the same in both cases.

The statistical description of the dynamics of sufficiently ergodic systems is usually based on the assumption that the energy is the sole integral of motion, such that the equilibrium system is entirely
determined by its energy. If this would be the case for our system, the two quenches would lead to the same equilibrium state. However, the dynamics according to the integrable Lieb--Liniger Hamiltonian are strongly constrained by the conserved quantities other than the total energy. By performing two different quenches to the same final Hamiltonian and energy, we investigate the effects of integrability on the postquench evolution of the Lieb--Liniger system.

The conserved energy following the quench is the energy of the system at time $t=0^+$,
\begin{align}\label{eq:addedEgammai}
    E_{\gamma_0\to\gamma} &\equiv \langle \psi(0^+) | \hat{H}(\gamma) | \psi(0^+) \rangle \nonumber \\
    &= E_{\rm{G}}(\gamma_0) +  (\gamma-\gamma_0) \frac{d E_{\mathrm{G}}(\gamma)}{ d\gamma}\Big|_{\gamma_0}
\end{align}
where $E_{\rm{G}}(\gamma_0)$ is the energy of the ground state $|\psi_0\rangle$ of the initial Hamiltonian $\hat{H}(\gamma_0)$ and we used the well-known result $g^{(2)}_{\gamma}(0) = n^{-2}N^{-1} d E_{\mathrm{G}}(\gamma) / d\gamma$~\cite{Gangardt2003a}, which implies that $E_{\gamma_0\to\gamma}$ is given by following the tangent to the curve $E_{\mathrm{G}}(\gamma)$ at $\gamma_0$ out to $\gamma$.
Here, $g^{(2)}_{\gamma_0}(0)\equiv \langle \psi_0 |\hat{g}^{(2)}(0) |\psi_0\rangle$ is the local second-order coherence in the initial state. 
In the case of a quench from the noninteracting ground state ($\gamma_0=0$), Eq.~\eqref{eq:addedEgammai} reduces to the simple expression $E_{0\to\gamma} = (N-1)n^2\gamma$~\cite{Muth2010a,Zill2015a}, implying that the energy imparted to the system during the quench diverges as $\gamma \rightarrow \infty$~\cite{Kormos2014}.  By contrast, in a quench from the Tonks--Girardeau limit $\gamma_0\to\infty$ to a finite interaction strength $\gamma$ the final energy is bounded from above, $E_{\infty \to \gamma} \leq E_{\mathrm{G}}(\infty)$, by the ground-state energy of the Tonks--Girardeau gas. 
Nevertheless, according to Eq.~(\ref{eq:addedEgammai}), a final interaction strength $0 < \gamma^* < 100$ such that $E_{100 \to \gamma^*} = E_{0 \to \gamma^*}$ does exist.

Here, we consider quenches of $N=5$ particles, and determine this final interaction strength to machine precision, inferring a value $\gamma^*\!=\!3.7660\dots$ from numerical solutions for the energy and local second-order coherence of the ground state at finite $\gamma$ (Sec.~\ref{subsec:SecondOrder}). 
We note that although the overlaps $C_{\{\lambda_j\}}$ of the initial state $|\psi_0\rangle$ with the eigenstates of $\hat{H}(\gamma^*)$ can be calculated analytically in the case of the quench from $\gamma_0=0$~\cite{Brockmann2014a,Brockmann2014b,Brockmann2014c}, for the quench from $\gamma_0=100$ no closed-form expressions for these quantities are known, and thus their numerical values must be determined using the semi-analytical methodology described in Sec.~\ref{subsec:correlation_functions}.

An important summary of the postquench expectation value of an operator [Eq.~\eqref{eq:operatorevolution}] is provided by the time-averaged value
\begin{equation}\label{eq:Otimeavg}
\overline{O} = \lim_{\tau \to \infty} \frac{1}{\tau} \int_{0}^{\tau} dt \, \langle \psi(t) | \hat{O} | \psi(t)\rangle .
\end{equation}
Neglecting degeneracies in the spectrum of $\hat{H}(\gamma^*)$ (see discussion in Appendix~\ref{app:DE}), such averages are given by the expectation values $\langle \hat{O} \rangle_{\mathrm{DE}} = \mathrm{Tr}\{\hat{\rho}_{\mathrm{DE}} \hat{O}\}$ of operators $\hat{O}$ in the density matrix
\begin{equation}\label{eq:rhoDE}
\hat{\rho}_{\mathrm{DE}} = \sum_{\{\lambda_j\}} | C_{\{\lambda_j\}} |^2 | \{ \lambda_j \} \rangle \langle \{\lambda_j \} |
\end{equation}
of the diagonal ensemble~\cite{Rigol2008,Kaminishi2014b}.

Formally, the sums in Eq.~\eqref{eq:psievolution},\eqref{eq:operatorevolution}, and \eqref{eq:rhoDE} range over an infinite number of eigenstates $|\{\lambda_j\}\rangle$, and thus the basis over which $|\psi(t)\rangle$ is expanded must be truncated in our numerical calculations.  By only including eigenstates with an absolute initial-state overlap $|C_{\{\lambda_j\}}|$ larger than some threshold, we consistently neglect small contributions to correlation functions from weakly occupied eigenstates and minimize the truncation error for a given basis size.  
We quantify this truncation error by the violations of the normalization and energy sum rules, as we discuss in Appendix~\ref{app:cutoff}.

\subsection{Evolution of two-body correlations}\label{subsec:localsecondorder}
%%%
\begin{figure}[t]
    \includegraphics[width=0.47\textwidth]{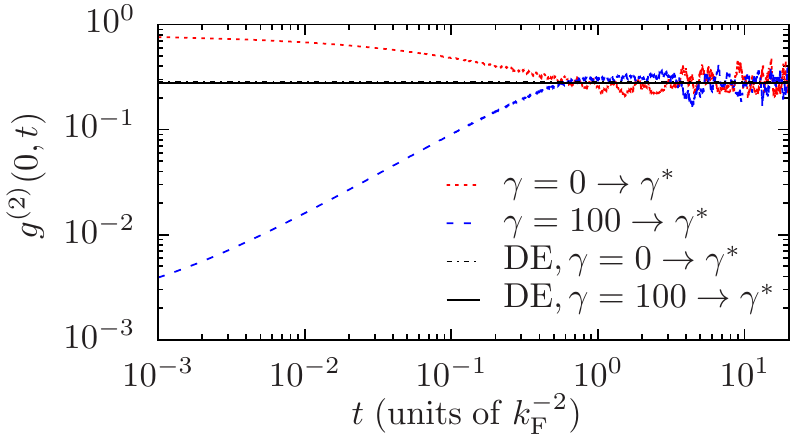}
    \caption{\label{fig:dyng2local} (Color online)  Time evolution of local second-order correlations for $N=5$ particles following quenches of the interaction strength to a final value $\gamma^*\!=\!3.7660\dots$ from initial values $\gamma_0=0$ (red dotted line) and $\gamma_0=100$ (blue dashed line). The horizontal solid (dot-dashed) line indicates the prediction of the diagonal ensemble for $g^{(2)}(0)$ for the quench from $\gamma_0 = 100$ ($\gamma_0 = 0$).}
\end{figure}
%%%
%%%
\begin{figure*}[t]
    \includegraphics[width=0.97\textwidth]{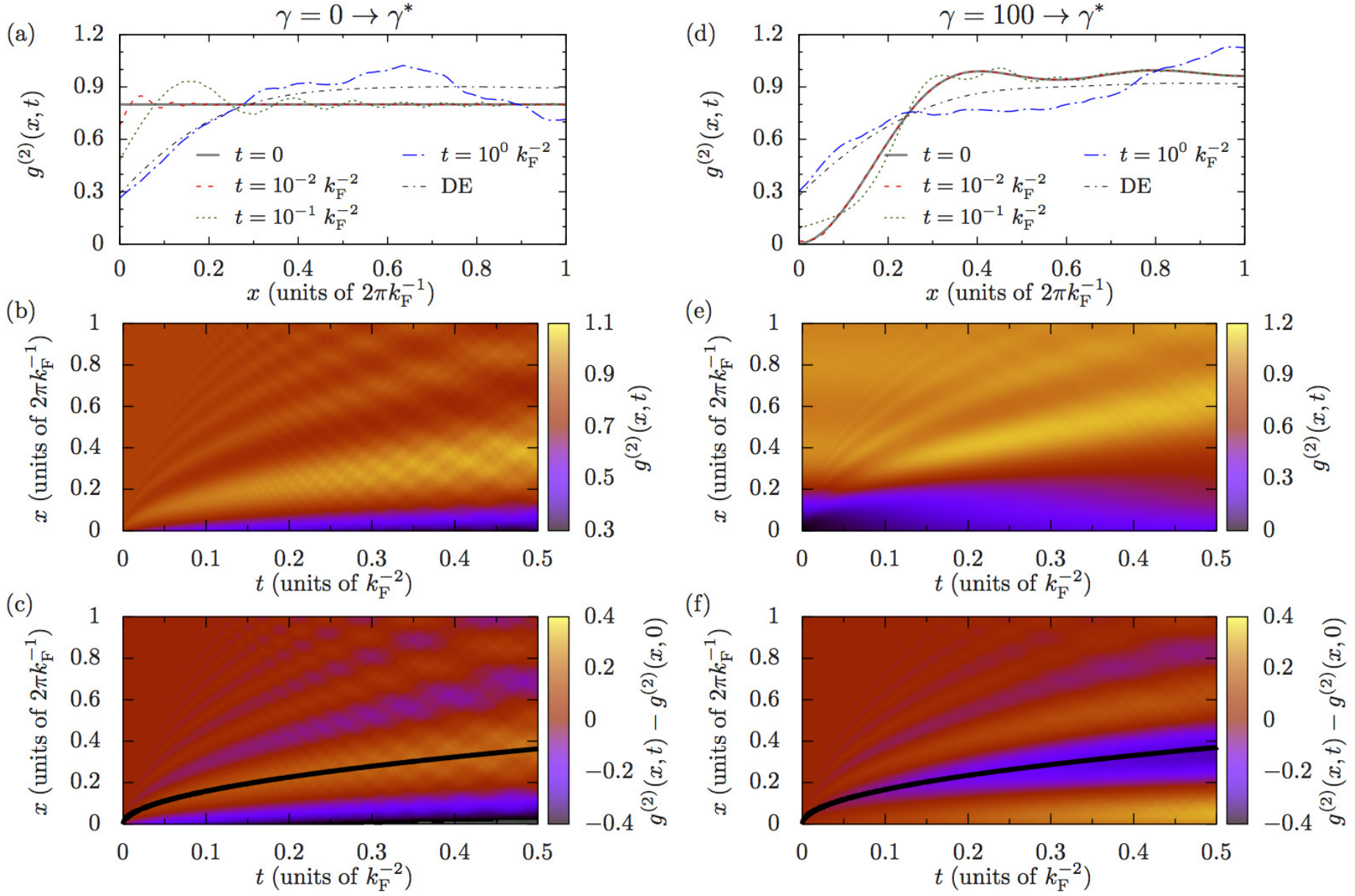}
    \caption{(Color online) Time evolution of the nonlocal second-order coherence function $g^{(2)}(x,t)$ following quenches of the interaction strength to $\gamma^*$ from initial values [(a),(b),(c)] $\gamma_0=0$ and [(d),(e),(f)] $\gamma_0=100$. All data is for $N=5$ particles. [(a),(d)] Correlation function $g^{(2)}(x,t)$ at four representative times $t$. Black dot-dashed lines indicate the predictions of the diagonal ensemble for the equilibrium form of this function. [(b),(e)] Evolution of coherence $g^{(2)}(x,t)$ and [(c),(f)] change in coherence $g^{(2)}(x,t) - g^{(2)}(x, t=0)$ for short times $t \leq 0.5k_\mathrm{{F}}^{-2}$.  Black lines in (c) and (f) indicate power-law fits to the position $x(t)$ of the first extremum of the correlation wave, which yield $x \propto t^{0.516\pm0.012}$ and $x \propto t^{ 0.496\pm 0.005}$ for quenches from $\gamma_0=0$ and $\gamma_0=100$, respectively.}\label{fig:g2dyn}
\end{figure*}
%%%

In Fig.~\ref{fig:dyng2local} we plot the time evolution of the local second-order coherence $g^{(2)}(0,t)$ for $N=5$ particles following quenches of the interaction strength from initial values $\gamma_0=0$ (red dotted line) and $\gamma_0=100$ (blue dashed line) to the common final value $\gamma^*$. 
For the quench from the noninteracting initial state ($\gamma_0=0$), as time evolves the local second-order coherence decays from its initial value $g^{(2)}(0,t = 0) = 1-N^{-1}$ before settling down to fluctuate about the diagonal-ensemble expectation value $g^{(2)}_{\mathrm{DE}}(0)$ (horizontal dot-dashed line). This behavior is consistent with results obtained for similar quenches of the interaction strength from zero to a positive value in Ref.~\cite{Zill2015a}.
For the quench from $\gamma_0=100$, the value of $g^{(2)}(0)$ in the initial ``fermionized'' state is $g^{(2)}(0) \approx 10^{-3}$. In this case $g^{(2)}(0,t)$ rises as time progresses, and then exhibits somewhat irregular oscillations about $g^{(2)}_{\mathrm{DE}}(0)$ (horizontal solid line).  We observe that the decay (growth) of $g^{(2)}(0,t)$ to its diagonal-ensemble value and the onset of irregular oscillations about this value occur on comparable time scales in the two quenches. 

We note that the predictions of the diagonal ensemble for the local second-order coherence $g^{(2)}_{\mathrm{DE}}(0)$ are very similar for the two quenches, despite the significant difference between the values of $g^{(2)}(0)$ in the two initial states.  However, they are clearly distinct --- $g^{(2)}_{\mathrm{DE}}(0)$ for the quench from the noninteracting state is in fact larger than that for the quench from the correlated state by an amount $\approx\!0.0125$, demonstrating that the system retains some memory of its initial state in the long time limit as is expected for an integrable system.  We analyze this difference in more detail in Sec.~\ref{subsec:contrib}.

We now turn our attention to the time evolution of the full non-local second-order correlation function $g^{(2)}(x,t)$. In Fig.~\ref{fig:g2dyn}(a) we show the dependence of $g^{(2)}(x,t)$ on separation $x$ for the quench from the noninteracting initial state at four representative times. [Note that the upper limit $x=2\pi k_\mathrm{{F}}^{-1}$ of the $x$ axis in Fig.~\ref{fig:g2dyn}(a) corresponds to $x=L/2$ in the present case of $N=5$ particles.] 
At $t=0$ (horizontal solid line), the second-order coherence has the constant form of the noninteracting ground state. At short times (e.g., $t=0.01 \, k_\mathrm{{F}}^{-2}$, red dashed line) a minimum in $g^{(2)}(x)$ develops at zero separation, together with the corresponding maximum required by the conservation of $\int_{0}^{L} dx \; g^{(2)}(x,t)$~\cite{Muth2010a}.  As time progresses a wave pattern of maxima and minima develops and propagates away from the origin (e.g., $t=0.1 \, k_\mathrm{{F}}^{-2}$, green dotted line). 
By time $t=1 \, k_\mathrm{{F}}^{-2}$ (blue dot-dashed line), the distinct maxima and minima of $g^{(2)}(x,t)$ have broadened in such a way that they are no longer clearly distinguishable and the correlation function agrees reasonably well with its diagonal-ensemble form (black dot-dashed line) for small separations $x \lesssim 0.25 \times 2\pi k_\mathrm{{F}}^{-1}$. 
In Fig.~\ref{fig:g2dyn}(b) we show the full space and time dependence of $g^{(2)}(x,t)$ following a quench from $\gamma_0=0$, which gives a more complete picture of the development of a correlation wave at short length scales and its propagation to larger values of $x$ as time progresses. The correlation wave we observe here is consistent with the results of previous investigations of the dynamics following the sudden introduction of repulsive interactions in an initially noninteracting gas~\cite{Gritsev2010,Kormos2014,Deuar2006,Muth2010a,Deuar2013}.

In Fig.~\ref{fig:g2dyn}(d) we plot the spatial form of $g^{(2)}(x,t)$ for the quench from $\gamma_0=100$ at the same four representative times considered in Fig.~\ref{fig:g2dyn}(a).  Despite the obvious distinction that the initial ($t=0$, solid grey line) correlation function is in the fermionized regime with $g^{(2)}(0)\ll1$, the behavior of $g^{(2)}(x,t)$ for this quench is qualitatively similar to that observed for the quench from $\gamma_0=0$, in that at early times (e.g., $t=0.01 k_\mathrm{{F}}^{-2}$, red dashed line), deviations from $g^{(2)}(x,t=0)$ occur only at small separations $x\ll 2\pi k_\mathrm{{F}}^{-1}$.
Moreover, as time evolves and $g^{(2)}(0,t)$ increases towards $g^{(2)}_{\mathrm{DE}}(0)$, larger modulations of $g^{(2)}(x,t)$ about its initial functional form develop (e.g., $t=0.1 k_\mathrm{{F}}^{-2}$, green dotted line). At later times (e.g., $t=1 k_\mathrm{{F}}^{-2}$, blue dot-dashed line), $g^{(2)}(x,t)$ is close to $g^{(2)}_{\mathrm{DE}}(x)$ at small separations $x \lesssim 0.25 \times 2\pi k_\mathrm{{F}}^{-1}$, but exhibits large excursions away from it at larger $x$.
In Fig.~\ref{fig:g2dyn}(e) we plot the full space and time dependence of $g^{(2)}(x,t)$ following the quench from $\gamma_0=100$. Although the behavior of $g^{(2)}(x,t)$ here obviously differs from that following a quench from the noninteracting initial state [Fig.~\ref{fig:g2dyn}(b)], with the ``fermionic'' depression around $x=0$ lessening rather than growing in magnitude, a similar pattern of propagating correlation waves in $g^{(2)}(x,t)$ can again be seen.

The correlation-wave pattern common to both quenches is more clearly exhibited by the \emph{change} $g^{(2)}(x,t) - g^{(2)}(x,0)$ in the correlation function following the quench, which we plot in Figs.~\ref{fig:g2dyn}(c) and~\ref{fig:g2dyn}(f).  This representation of the postquench second-order coherence of the system reveals a remarkably similar pattern of propagating waves in both cases, although the maxima and minima of the two wave patterns are inverted relative to one another. 
Fitting a power law to the position $x(t)$ of the first propagating extremum of each of the two correlation waves, we find $x \propto t^{0.516\pm0.012}$ for the quench from $\gamma_0=0$ and $x \propto t^{0.496 \pm 0.005}$ for the quench from  $\gamma_0=100$, which we indicate by the solid black lines in Figs.~\ref{fig:g2dyn}(c)~and~\ref{fig:g2dyn}(f). These power-law trajectories are consistent with the ``telescoping'' $x \propto t^{1/2}$ behavior obtained for a quench $\gamma=0 \rightarrow \infty$ in Ref.~\cite{Kormos2014}, and for quenches from finite repulsive interactions to the noninteracting limit in Ref.~\cite{Imambekov2009} (see also Ref.~\cite{Mossel2012a}).
The small scale features on top of the main propagating extrema differ for the two quenches, with fast oscillations appearing more pronounced for the quench $\gamma=0\rightarrow \gamma^*$ in Fig.~\ref{fig:g2dyn}(c). Even though hardly visible in Fig.~\ref{fig:g2dyn}(f), they are still present for the quench from $\gamma=100 \rightarrow \gamma^*$, but due to the different distribution of overlaps in the final basis compared to the quench from $\gamma_0=0$ (cf.~Sec.~\ref{subsec:contrib}), they contain more high-frequency components and therefore the fine structure differs.

\subsection{Time-averaged correlations}\label{subsec:relaxed}
%%%
\begin{figure}[t]
    \includegraphics[width=0.48\textwidth]{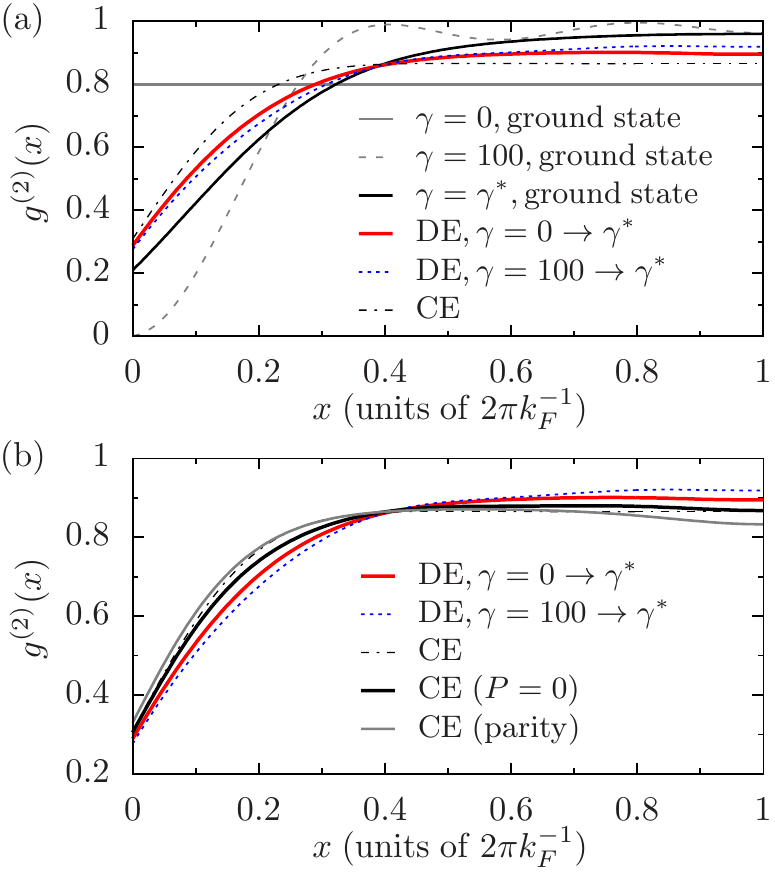}
    \caption{\label{fig:g2dynstat} (Color online) Time-averaged second-order correlation functions following quenches of the interaction strength to $\gamma^*\!=\!3.7660\dots$ from initial values $\gamma_0=0$ (red solid line) and $\gamma_0=100$ (blue dotted line). Results are for $N=5$ particles.  
(a) The correlation functions $g^{(2)}(x)$ in the initial states with $\gamma_0=0$ (horizontal solid line) and $\gamma_0=100$ (grey dashed line), as well as for the ground state at $\gamma=\gamma^*$ (solid black line) are also indicated for comparison.  The black dot-dashed line corresponds to the thermal value of the correlation function following relaxation, as predicted by the canonical ensemble (see text).
(b) Comparison of the time-averaged second order correlation functions to the various ensembles defined in the text: The standard canonical ensemble (black dot-dashed line), the canonical ensemble restricted to zero-momentum eigenstates (black solid line), and the canonical ensemble restricted to parity-invariant states (grey solid line). 
}
\end{figure}
%%%
We now compare the time-averaged second-order correlation functions following the two quenches with the form of this function that would be obtained if, following the quench, the system relaxed to thermal equilibrium. 
As in Ref.~\cite{Zill2015a} we make use of the canonical ensemble, for which the density matrix is given by
\begin{equation} \label{eq:rhoCE}
\hat{\rho}_{\mathrm{CE}} = Z_{\mathrm{CE}}^{-1} \sum_{ \{ \lambda_j \}} \mathrm{e}^{-\beta E_{\{\lambda_j\}}} \; | \{\lambda_j \} \rangle \langle \{ \lambda_j \} |,
\end{equation}
where the partition function $Z_{\mathrm{CE}} = \sum_{ \{ \lambda_j \}} \exp(-\beta E_{\{\lambda_j\}})$. The inverse temperature $\beta$ is determined implicitly by fixing the mean energy in the state $\hat{\rho}_\mathrm{CE}$ to the common postquench energy, i.e., $\mathrm{Tr}\{ \hat{\rho}_{\mathrm{CE}} \hat{H}(\gamma^*)\} = E_{0\to\gamma^*}$. 
The sum in Eq.~\eqref{eq:rhoCE}, like that in Eq.~\eqref{eq:rhoDE}, formally ranges over an infinite number of eigenstates.  We therefore truncate this sum by applying a cutoff in energy, as described in Appendix~\ref{app:cutoff}.

In Fig.~\ref{fig:g2dynstat}(a) we plot the second-order correlation function $g^{(2)}_{\mathrm{CE}}(x) = \mathrm{Tr}\{\hat{\rho}_{\mathrm{CE}}\,\hat{g}^{(2)}(0,x)\}$ in the canonical ensemble (black dot-dashed line), along with the diagonal-ensemble predictions $g^{(2)}_{\mathrm{DE}}(x)$ for the quenches from $\gamma_0=0$ (red solid line) and from $\gamma_0=100$ (blue dotted line). For comparison we also plot the correlation functions in the initial states with $\gamma_0=0$ (horizontal line), $\gamma_0=100$ (grey dashed line), as well as the ground state for $\gamma=\gamma^*$ (solid black line).  
For the quench from $\gamma_0=0$, the time-averaged value $g^{(2)}_{\mathrm{DE}}(0)$ is smaller than the corresponding thermal value $g^{(2)}_{\mathrm{CE}}(0)$, consistent with the results of Refs.~\cite{DeNardis2014,Kormos2013,Zill2015a}.  In fact $g^{(2)}_{\mathrm{DE}}(x)$ is suppressed below $g^{(2)}_{\mathrm{CE}}(x)$ over a range of separations $x \lesssim 0.4 \times 2\pi  k_\mathrm{{F}}^{-1}$. Correspondingly, $g^{(2)}_{\mathrm{DE}}(x)>g^{(2)}_{\mathrm{CE}}(x)$ at larger separations $x$ due to particle number and momentum conservation.
For the quench $\gamma=100\to\gamma^*$, the diagonal-ensemble coherence function $g^{(2)}_{\mathrm{DE}}(x)$ is similar in shape to that of the quench from $\gamma_0=0$. However, it is somewhat smaller at $x=0$, and correspondingly larger at large $x$. This indicates some memory of the initial state preserved by the dynamics of the integrable Lieb--Liniger system~\cite{Mussardo2013,Rigol2007}.
Despite these differences, on the whole both functions $g^{(2)}_{\mathrm{DE}}(x)$ are comparable to $g^{(2)}_{\mathrm{CE}}(x)$ (cf.~also Ref.~\cite{Muth2010a}).  
We note, however, that they are also both reasonably close to the ground state result for $g^{(2)}(x)$ at interaction strength $\gamma^*$ (solid black line), although the local value $g^{(2)}_{\mathrm{DE}}(0)$ for both quenches is much closer to the thermal value than the ground state value.

Since the system is in its ground state before the quench for both $\gamma_0=0$ and $\gamma_0=100$, and the total momentum operator $\hat{P}$ commutes with the Hamiltonian, the postquench states at $\gamma^*$ only have support on eigenstates with total momentum $P=0$. Furthermore, the spatially structureless initial state at $\gamma_0=0$ implies additional parity-invariance ($\{\lambda_j\} = \{-\lambda_j\}$) in Bethe rapidity space for the postquench eigenstates~\cite{Brockmann2014a,Brockmann2014b,Brockmann2014c}.  Thus an interesting question to ask is if we constructed a canonical density matrix~\eqref{eq:rhoCE} restricted to $P=0$ states, or one further restricted to  parity-invariant states (which are a subset of the $P=0$ states), would these yield better agreement with the diagonal ensemble predictions for the quenches?   We have performed these constructions with the temperature in both cases fixed via the postquench energy in the same way as for the canonical ensemble, cf.~Eq.~\eqref{eq:rhoCE} and the following text. 

In Fig.~\ref{fig:g2dynstat}(b), we plot the resulting second-order correlation function $g^{(2)}_{\mathrm{CE}}(x) = \mathrm{Tr}\{\hat{\rho}_{\mathrm{CE}}\,\hat{g}^{(2)}(0,x)\}$ for the standard canonical ensemble (black dot-dashed line), as well as in the restricted $P=0$ ensemble (solid black line), and the parity-invariant ensemble (solid grey line).
We also include the diagonal-ensemble predictions $g^{(2)}_{\mathrm{DE}}(x)$ for the quenches from $\gamma_0=0$ (red solid line) and from $\gamma_0=100$ (blue dotted line).  It can be seen that the restricted ensembles give results for the correlation function that are quite close to the standard canonical ensemble, and are no closer to the diagonal ensemble results.

\subsection{Contributions to relaxed correlation functions}\label{subsec:contrib}
%%%
\begin{figure}[t]
    \includegraphics[width=0.48\textwidth]{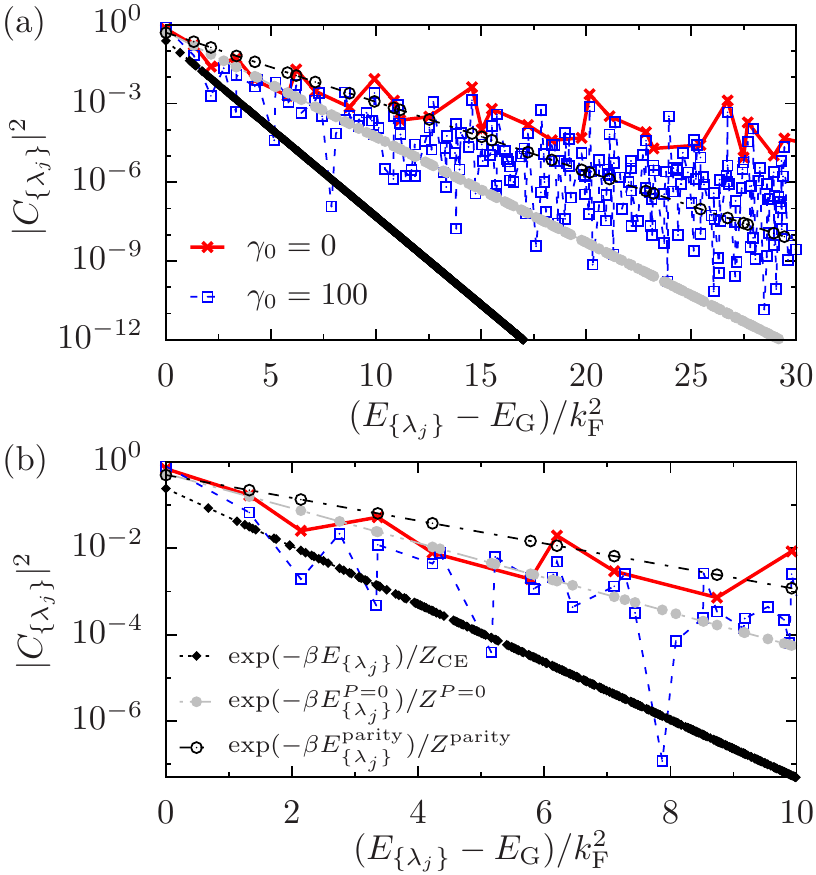}
    \caption{\label{fig:overlapdist} (Color online) (a) Populations $|C_{\{\lambda_j\}}|^2$ of eigenstates with energies $E_{\{\lambda_j\}}$ following quenches to $\gamma^*=\!3.7660\dots$ from $\gamma_0=0$ (red crosses) and $\gamma_0=100$ (blue squares).  Note that the $y$-axis is plotted on a logarithmic scale. For the quench from $\gamma_0=100$, additional non-parity-invariant states appear in degenerate, parity-conjugate pairs and since their contributions is identical, the points lie on top of each other. 
The black dotted line with filled black circles represents the populations $\exp(-\beta E_{\{\lambda_j\}})/Z_{\mathrm{CE}}$ of eigenstates with energies $E_{\{\lambda_j\}}$ for the canonical ensemble.  The grey line with grey filled circles, and the black dashed line with empty black circles are the corresponding results for the $P=0$ restricted ensemble, and the parity-restricted ensemble, respectively. 
(b) Low-energy part of (a).
}
\end{figure}
%%%
The relaxation of the nonlocal correlations $g^{(2)}(x,t)$ takes place on a similar time scale to that of the local coherence $g^{(2)}(0,t)$ for both of the quenches considered here.  This should be contrasted with, e.g., the behavior following a quench from the noninteracting limit to $\gamma=100$ reported in Ref.~\cite{Zill2015a}, in which $g^{(2)}(0,t)$ decays rapidly and the development and propagation of correlation waves occurs over a significantly longer time scale.  We identify the absence of a significant separation of the time scales of local and nonlocal evolution here as a consequence of the fact that only a small number of eigenstates contribute significantly to the postquench dynamics (cf.\ Ref.~\cite{Zill2015a} and references therein). Indeed, we find that the purity $\Gamma_{\mathrm{DE}} \equiv \mathrm{Tr}\{ (\hat{\rho}_{\mathrm{DE}})^2 \}$ of the diagonal-ensemble density matrix takes values $\approx\!0.52$ for the quench $\gamma=0\to\gamma^*$ and $\approx\!0.63$ for the quench $\gamma=100\to\gamma^*$, indicating rather weak participation of the eigenstates $|\{\lambda_j\}\rangle$ in the dynamics.   The difference in the purities can largely be attributed to the somewhat greater occupation of the ground state of $\hat{H}(\gamma^*)$ following the quench from the $\gamma=100$ initial state. 

To further illustrate the difference in the final states, in Fig.~\ref{fig:overlapdist} we plot the occupations of eigenstates with energy $E_{\{\lambda_j\}}$ for the quenches from $\gamma_0=0$ (red crosses) and $\gamma_0=100$ (blue squares). For the quench from $\gamma_0=100$, significantly more eigenstates have occupations above a given threshold than in the case of $\gamma_0=0$, resulting in a much larger basis size in this case.  However, the occupation of the ground state of $\hat{H}(\gamma^*)$ is somewhat larger for the quench from $\gamma_0=100$ than for  $\gamma_0=0$, and the low-lying excited states are comparatively weakly occupied for $\gamma_0=100$, cf.~Fig.~\ref{fig:overlapdist}(b).  This result is reasonably intuitive, as the ground state for $\gamma = \gamma^*$ is moderately correlated, and will be more similar to the $\gamma=100$  than the $\gamma=0$ ground state.  The distribution of normalization over eigenstates $|\{\lambda_j\}\rangle$ is thus more sharply ``localized'' on the ground state in this case, resulting in the somewhat larger value of the purity $\Gamma_\mathrm{DE}$ following this quench.

For comparison, we also plot the occupations of the three ensembles introduced in Sec.~\ref{subsec:relaxed} in Fig.~\ref{fig:overlapdist}. The restrictions lead to a reduction in available eigenstates for any given energy-window, and correspondingly the temperature of the canonical ensemble is smaller than that of the $P=0$ ensemble, which is in turn smaller than that of the parity-invariant ensemble.
The occupations of eigenstates for the quench from $\gamma_0=0$ (red crosses) and from $\gamma_0=100$ (blue squares) are suggestive of power-law decay at high energies. For small energies on the other hand, Fig.~\ref{fig:overlapdist}(b) shows that the functional form is not incompatible with exponential decay.

\section{Conclusions}\label{sec:conclusions}
We have described a method to calculate matrix elements between eigenstates of the Lieb--Liniger model of one-dimensional delta-interacting bosons. This method is based on the coordinate Bethe ansatz, which generates a complete set of energy eigenfunctions for any fixed coupling strength. This allows us to obtain overlaps between eigenstates of different Hamiltonians, as well as expressions for correlation functions. By introducing periodic boundary conditions, we obtained expressions amenable to numerical evaluation. 
We applied our methodology to the evaluation of first-, second-, third-, and fourth-order correlation functions in the ground state of the Lieb--Liniger model for various values of the interparticle interaction strength.  Our results indicate that although the correlations of the system are in general distorted by the small system size, finite-size effects become increasingly less significant with increasing interaction strength and decreasing spatial separation.

Out of equilibrium, we investigated the dynamics of relaxation after a quantum quench of the interparticle interaction strength towards a non-thermal steady state. Starting from two different initial states, we quenched to a common final interaction strength $\gamma^*$ chosen in such a way that both postquench energies were the same. 
Our calculations reveal a similar relaxation process for the second-order coherence $g^{(2)}(x,t)$ for both initial states: the build-up of correlations on short interparticle distances and their propagation through the system as time progresses.
The time-averaged second-order correlation functions in both cases disagreed with the prediction for thermal equilibrium and were biased, relative to one another, towards their pre-quench forms --- an intuitive result given the integrability of the system.  In the future it would be interesting to study quenches from other initial states with the same final energy to explore how the memory of the initial state is manifest in different situations.

Although our method is restricted to small system sizes due to computational complexity and here only applied to five particles out of equilibrium, we were able to obtain the dynamical evolution as well as time-averaged correlation functions to high precision.
Finally we note that the evaluation of matrix elements of the Lieb--Liniger model with this method is not restricted to real-valued Bethe rapidities, opening the door to investigating the nonequilibrium dynamics of attractively interacting systems (where the rapidities become complex-valued) and that following quenches from more complex initial states.

\begin{acknowledgments}
This work was supported by ARC Discovery Project, Grant No.\ DP110101047 (J.C.Z., T.M.W., K.V.K., and M.J.D.) and by the EU-FET Proactive grant AQuS, Project No. 640800 (TG). M.J.D. acknowledges the support of the JILA Visiting Fellows program.
\end{acknowledgments}

\appendix

\section{Basis-set truncation}\label{app:cutoff}
The Hilbert space of the Lieb--Liniger model is infinite dimensional, and therefore the sums in Eqs.~\eqref{eq:psievolution},~\eqref{eq:operatorevolution},~and~\eqref{eq:rhoDE} must be truncated for numerical purposes.
Here, we  provide details of the truncation scheme for the two different initial states we considered in Sec.~\ref{sec:Dynamics}, and explain how we quantify the error resulting from this truncation.

For the quench from $\gamma_0=0$, the initial state $|\psi_0\rangle$ only has nonzero overlap with eigenstates $| \{\lambda_j \}\rangle$ of $\hat{H}(\gamma^*)$ that are parity invariant (i.e., eigenstates for which $\{\lambda_j\} = \{-\lambda_j\}$) and, {\emph{a fortiori}}, have zero total momentum $P$~\cite{Calabrese2014}. The strongly-correlated initial state of the quench from $\gamma_0=100$ similarly has zero overlap with eigenstates $| \{\lambda_j \}\rangle$ with nonzero total momentum, but in this case states contributing to  $|\psi(t)\rangle$, and thus $\hat{\rho}_{\mathrm{DE}}$, need not be parity-invariant in general. 
For $\gamma_0=0$ our results for the overlaps agree with recently obtained analytical expressions~\cite{Brockmann2014c,Brockmann2014b}, which predict real positive overlaps, given the phase convention implicit in Eq.~\eqref{eq:eigenfunction_unordered}, for quenches to $\gamma > 0$. For $\gamma_0=100$, we find that the overlaps are still real, but are no longer restricted to positive values.

We briefly summarize our procedure to determine the cutoff here --- see Appendix A of Ref.~\cite{Zill2015a} for an extended discussion for the case of parity-invariant states. It can be shown~\cite{Yang1969} that the solutions $\{\lambda_j\}$ of the Bethe equations~\eqref{eq:Betheeq} are in one-to-one correspondence with the numbers $m_j$ that appear in Eq.~\eqref{eq:Betheeq}. This allows us to uniquely label states by the set $\{m_j\}$.  Without loss of generality, we order the numbers $m_j$ such that $m_1>m_2>\cdots>m_{N-1}>m_N$, and we only need consider states for which $\sum_j m_j=0$, corresponding to zero total momentum $P$. 
We specialize hereafter to the case $N=5$, which is the largest $N$ for which we consider the dynamics in this article.  
The states can be grouped into families, labelled by $m_1$.  We have found empirically that within each such family, the eigenstate $(m_1,1,0,-1,-m_1)$ has the largest absolute overlap $|\langle \{\lambda_j\} | \psi_0 \rangle|$ with the initial state, for both initial states we consider ($\gamma_0=0$ and $\gamma_0=100$).  Furthermore, this overlap is larger than that of the most significantly contributing eigenstate $(m_1+1,1,0,-1,-m_1-1)$ of the following family ($m_1+1$).  We therefore construct the basis by considering in turn each family $m_1$ and including all states within that family for which the overlap with the initial state exceeds our chosen threshold value $C_\mathrm{min}$.  Eventually, for some value of $m_1$, even the eigenstate $(m_1,1,0,-1,-m_1)$ has overlap with $|\psi_0\rangle$ smaller than the threshold, at which point all states that meet the threshold have been accounted for. 

We note that the Lieb--Liniger model has an infinite number of conserved charges $[\hat{Q}^{(m)},\hat{H}(\gamma)]=0$; $m=0,1,2,\dots$, with eigenvalues given by $\hat{Q}^{(m)} | \{ \lambda_j \} \rangle = \sum_{l=1}^{N} \lambda_l^m | \{ \lambda_j \} \rangle$. However, for a quench from $\gamma_0=0$ their expectation values in the diagonal ensemble $\langle \hat{Q}^{(m)} \rangle_\mathrm{DE}$ diverge for all even $m \geq 4$ \cite{Brockmann2014c,Brockmann2014b}. Our numerical results suggest that this is also the case for quenches from $\gamma_0>0$ (indeed, they diverge for almost all states but eigenstates~\cite{Davies1990,Davies2011}).
For all odd values $m$, the expectation values of the corresponding conserved charges $\hat{Q}^{(m)}$ are identically zero for our initial states and quench protocol. Thus, the only nontrivial and regular conserved quantities are the particle number ($m=0$) and energy ($m=2$). 
As in Ref.~\cite{Zill2015a}, we quantify the saturation of the normalization and energy sum rules by the sum-rule violations 
\begin{align}
\Delta N &= 1 - \sum_{\{\lambda_j\}} |C_{\{\lambda_j\}}|^2 , \\ 
 \Delta E &= 1 -  \frac{1}{E_{\gamma_0\to\gamma}} \sum_{\{\lambda_j\}} |C_{\{\lambda_j\}}|^2 \sum_{l=1}^{N} (\lambda_l)^2,
\end{align} 
respectively, where $E_{\gamma_0\to\gamma}$ is the exact postquench energy [Eq.~\eqref{eq:addedEgammai}]. We note that the calculation of time-dependent observables involves a double sum over $\{\lambda_j\}$, and is therefore more numerically demanding than the calculation of expectation values in the DE.  Moreover, the calculation of the local coherence $g^{(2)}(0,t)$ is much less demanding than that of the full nonlocal $g^{(2)}(x,t)$.  We therefore use different thresholds $C_\mathrm{min}$, resulting in different basis sizes and sum-rule violations, in the calculation of $g^{(2)}(0,t)$, $g^{(2)}(x,t)$, and $g^{(2)}_\mathrm{DE}(x)$, as indicated in Table~\ref{tab:cutoffs}. 
We note that the energy sum rule is in general less well satisfied than the normalization sum rule, due to the $\propto\!\lambda^{-4}$ tail of the diagonal-ensemble distribution of eigenstates~\cite{Zill2015a}.  We find also that both sum rules are less well satisfied for the quench $\gamma=100\rightarrow \gamma^*$, despite the truncation procedure described above resulting in more than five times as many basis states being employed in its solution than are used in the quench $\gamma=0\rightarrow \gamma^*$. 

For expectation values in the CE [Eq.~\eqref{eq:rhoCE}], we truncate the basis by retaining all states with energies below some cutoff $E_\mathrm{cut}$.  The inverse temperature $\beta$ is then chosen to minimize the energy sum-rule violation $\Delta E$. The normalization sum rule is fulfilled by construction.  
Since all states (not only those with zero momentum) contribute to this sum, the number of eigenstates involved in canonical-ensemble calculations is much larger than that in diagonal-ensemble calculations. For the canonical-ensemble correlation function plotted in Fig.~\ref{fig:g2dynstat} we used an energy cutoff of $3.2 \times 10^2 \, k_\mathrm{{F}}^2$, which yields a basis of $2.1 \times 10^6$ eigenstates $|\{\lambda_j\}\rangle$. 
We checked that this cutoff is sufficiently large to ensure saturation of $g^{(2)}_\mathrm{CE}(x)$ (Fig.~\ref{fig:g2dynstat}).
For the ensemble restricted to $P=0$ eigenstates [Fig.~\ref{fig:g2dynstat}(b)], we used an energy cut-off of $6.4 \times 10^5 \, k_\mathrm{{F}}^2$, corresponding to $44530$ eigenstates, while for the parity-invariant ensemble we used an energy cut-off of $8.5 \times 10^6 \, k_\mathrm{{F}}^2$, corresponding to $64204$ eigenstates.

%%%
\begin{table}
    \centering
    \caption{\label{tab:cutoffs}  Basis-set sizes and sum-rule violations for full non-local, time-evolving second-order coherence $g^{(2)}(x,t)$, for local, time-evolving second-order coherence $g^{(2)}(x=0,t)$, and for time-averaged second-order coherence $g^{(2)}_{\mathrm{DE}}(x)$ following quenches from $\gamma_0=0$, and $\gamma_0=100$ to $\gamma^*=\!3.7660\dots$.}
    \begin{tabular}{cccccc}
    \hline \hline
\vspace{0.1cm}
    $\gamma_0 $ & $\;$ Type$^\mathrm{a}$ $\;$ & $C_{\mathrm{min}}$ & No. states & $\quad \Delta N \quad$ & $\quad\enskip \Delta E/k_\mathrm{{F}}^2 \quad\enskip$  \\
    \hline 
\vspace{0.1cm} 
   $0$ & $g^{(2)}(x,t)$ & $5\times 10^{-5}$ & $673$ & $7\times 10^{-7}$ & $6\times 10^{-3}$ \\  
\vspace{0.1cm}
    $0$ & $g^{(2)}(0,t)$ &  $1\times 10^{-5}$ & $1704$ & $7\times 10^{-8}$ & $3\times 10^{-3}$ \\
\vspace{0.1cm}
    $0$ & $g^{(2)}_{\mathrm{DE}}(x)$ &  $1 \times 10^{-6}$ & $6282$ & $2\times 10^{-9}$ & $8\times 10^{-4}$ \\
\vspace{0.1cm}
    $100$ & $g^{(2)}(x,t)$  &  $5\times 10^{-5}$ & $3704$ & $4\times 10^{-6}$ & $4\times 10^{-2}$  \\ 
\vspace{0.1cm}
    $100$ & $g^{(2)}(0,t)$ &  $1\times 10^{-5} $  & $10473$ & $5\times 10^{-7}$ & $3\times 10^{-2}$ \\
\vspace{0.1cm}
    $100$ & $g^{(2)}_{\mathrm{DE}}(x)$ &  $1\times 10^{-6}$ & $ 43918 $ & $ 2\times 10^{-8}$ & $2\times 10^{-3}$ \\
    \hline\hline
    \end{tabular}
    \flushleft
    $^\mathrm{a}$ Occupations of the $g^{(2)}_\mathrm{DE}(x)$ basis set are used in the calculation of $\Gamma_\mathrm{DE}$ (Sec.~\ref{subsec:contrib}).
\end{table}
%%%

\section{Time-averaged correlation functions and the diagonal ensemble}\label{app:DE}
%%%
\begin{figure}[t]
    \includegraphics[width=0.48\textwidth]{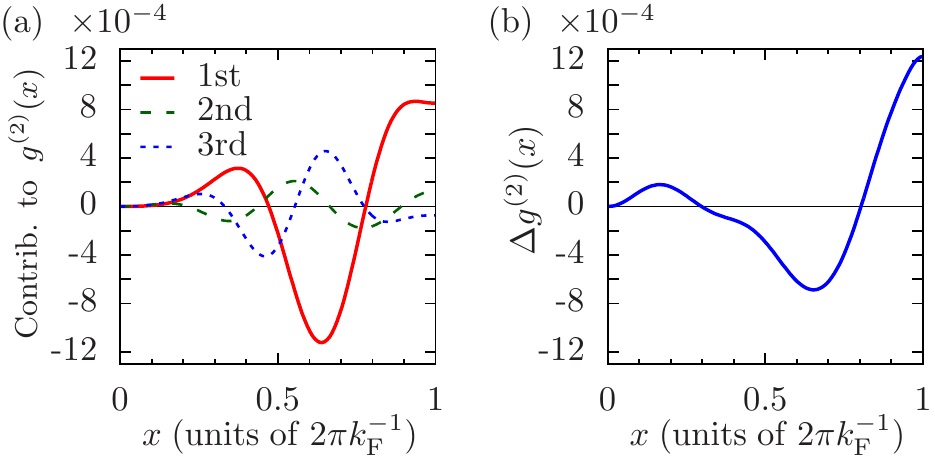}
    \caption{\label{fig:ensemblediff} (Color online) Contributions of degenerate energy eigenstates to the time-averaged second-order correlation function following a quench from $\gamma_0=100$ to $\gamma^*=\!3.7660\dots$ for $N=5$ particles.  (a) Contributions $C^{}_{\{\lambda_j\}}C^*_{\{-\lambda_j\}} \langle \{-\lambda_j\}|\hat{g}^{(2)}(0,x)|\{\lambda_j\}\rangle + \mathrm{c.c.}$ of off-diagonal matrix elements corresponding to the three largest weights $C^{}_{\{\lambda_j\}}C^*_{\{-\lambda_j\}}$. (b) Total contribution of degenerate energy eigenstates.
 }
\end{figure}
%%%
The time-averaged expectation value [Eq.~\eqref{eq:Otimeavg}] of an operator $\hat{O}$ can be expressed as an expectation $\overline{O}=\mathrm{Tr}\{\overline{\hat{\rho}}\,\hat{O}\}$ in the time-averaged density matrix
\begin{align}\label{eq:rhot}
\overline{\hat{\rho}} &\equiv \lim_{\tau \to \infty} \frac{1}{\tau} \int_{0}^{\tau} dt \, |\psi(t)\rangle\langle\psi(t)| \nonumber \\
&= \sum_{\{\lambda_j\}} | C_{\{\lambda_j\}} |^2 | \{ \lambda_j \} \rangle \langle \{\lambda_j \} | \nonumber \\
&\quad+\! \sum_{\{\lambda^{}_j\}\neq\{\lambda'_j\}} \delta_{E_{\{\lambda_j^{}\}},E_{\{\lambda'_j\}}} C^*_{\{\lambda'_j\}} C_{\{\lambda_j\}}  | \{ \lambda_j \} \rangle \langle \{\lambda'_j \} | \; .
\end{align}
The first term in Eq.~\eqref{eq:rhot} is simply the diagonal-ensemble density matrix $\hat{\rho}_\mathrm{DE}$ [Eq.~\eqref{eq:rhoDE}], to which $\overline{\hat{\rho}}$ reduces in the absence of degeneracies in the spectrum of $\hat{H}(\gamma)$. 
This is the case for the quench from $\gamma_0=0$, as the only eigenstates of $\hat{H}(\gamma^*)$ with nonvanishing overlaps with $|\psi_0\rangle$ in this case are the parity-invariant states $|\{\lambda_j\}\rangle$ with $\{\lambda_j\}$=$\{-\lambda_j\}$, which are nondegenerate (see Ref.~\cite{Zill2015a} and references therein).
By contrast, in a quench from $\gamma_0>0$, $|\psi(t)\rangle$ has support on non-parity-invariant states $|\{\lambda_j\}\rangle$, which are degenerate with their parity conjugates $|\{-\lambda_j\}\rangle$.

In general such degeneracies can have observable consequences for time-averaged expectation values~\cite{Kaminishi2014b}. However, as can be seen from Fig.~\ref{fig:ensemblediff}, the correction to $g^{(2)}_\mathrm{DE}(x)$ due to the contributions of degenerate eigenstates in the case of the quench from $\gamma_0=100$ is small. 
It is straightforward to show that the elements $\langle\{-\lambda_j\}|\hat{g}^{(2)}(0)|\{\lambda_j\}\rangle$ of the local second-order coherence between parity-conjugate states must vanish due to symmetry considerations.  
At larger separations $x$, the matrix elements between these pairs of states are nonzero, as illustrated in Fig.~\ref{fig:ensemblediff}(a).  However, these contributions are small compared to the diagonal-ensemble result $g^{(2)}_\mathrm{DE}(x)$, and indeed the total contribution of all parity-conjugate states in our finite-basis description [Fig.~\ref{fig:ensemblediff}(b)] would yield a barely visible correction to the function $g^{(2)}_\mathrm{DE}(x)$ plotted in Fig.~\ref{fig:g2dynstat}. We note also that the substitution of $\hat{\rho}_{\mathrm{DE}}$ for the time-averaged density matrix $\overline{\hat{\rho}}$ introduces negligible error in the calculation of the purity of this matrix (Sec.~\ref{subsec:contrib}).

\bibliographystyle{prsty}

\end{document}